\pdfoutput=1
\documentclass[letterpaper,10pt,conference]{ieeeconf}
\IEEEoverridecommandlockouts
\usepackage{amsmath,amssymb}
\usepackage{graphicx}
\usepackage{xcolor}
\usepackage{bm}
\usepackage{array}
\usepackage{cite}
\graphicspath{{./}{figs/}}
\usepackage{algorithm}
\usepackage{algpseudocode}
\usepackage{mathrsfs}
\let\IEEElabelindentsaved\labelindent\let\labelindent\relax
\usepackage{enumitem}
\let\labelindent\IEEElabelindentsaved
\usepackage[caption=false,font=footnotesize]{subfig}
\newcommand{\DKL}{\mathbb{D}_{\mathrm{KL}}}
\newcommand{\E}{\mathbb{E}}
\newcommand{\argmin}{\mathop{\rm argmin}\limits}
\newcommand{\Sel}{\mathcal{S}}
\newcommand{\ESS}{\mathrm{ESS}}

\newcommand{\figref}[1]{Fig.~\ref{#1}}
\newcommand{\tabref}[1]{Table~\ref{#1}}
\newcommand{\algorithmref}[1]{Algorithm~\ref{#1}}

\newlength{\figboxw}
\newcommand{\figorbox}[2]{\IfFileExists{#2}{\includegraphics[width=#1]{#2}}{\IfFileExists{figs/#2}{\includegraphics[width=#1]{figs/#2}}{\setlength{\figboxw}{#1}\framebox[\figboxw]{\rule{0pt}{0.6\figboxw}\scriptsize\detokenize{#2}}}}}

\newtheorem{assumption}{\bf{Assumption}}
\newtheorem{lemma}{\bf{Lemma}}
\newtheorem{theorem}{\bf{Theorem}}

\newtheorem{corollary}{\bf{Corollary}}
\newtheorem{remark}{Remark}
\newtheorem{Definition}{\bf{Definition}}
\newtheorem{Problem}{\bf{Problem}}
\renewenvironment{proof}{\noindent\textit{Proof:}}{\hspace*{\fill}~$\blacksquare$\par}
\newenvironment{proofof}[1]{\noindent\textit{Proof of #1:}}{\hspace*{\fill}~$\blacksquare$\par}

\title{\LARGE \bf
Laguerre-MPPI
}

\author{Yusuke~Tahori$^{1}$, Masaki~Inoue$^{1}$, Hayate~Irie$^{1}$, Tomohiro~Taki$^{2}$ and Takashi~Hatano$^{2}$%
\thanks{This work was supported by Mazda Motor Corporation.}%
\thanks{$^{1}$Yusuke~Tahori, Masaki~Inoue and Hayate~Irie are with the Department of Applied Physics and Physico-Informatics, Faculty of Science and Technology, Keio University, Kanagawa, Japan.\newline
       {\tt\small [tahori@keio.jp, minoue.z6@keio.jp, irihaya0913@keio.jp]}}%
\thanks{$^{2}$Tomohiro~Taki and Takashi~Hatano are with Mazda Motor Corporation,
        Hiroshima, Japan.\newline {\tt\small [taki.t@mazda.co.jp, hatano.taka@mazda.co.jp]}}%
}

\begin{document}
\maketitle
\thispagestyle{empty}
\pagestyle{empty}

\begin{abstract}
Model Predictive Path Integral control (MPPI) is a sampling-based control method that handles nonlinear dynamics and non-convex costs without gradient information.
Since MPPI samples the entire input sequence, longer prediction horizons lead to a higher-dimensional sampling space, potentially degrading control performance.
This paper proposes Laguerre-MPPI, which parameterizes the control input sequence using discrete-time Laguerre basis functions and samples the basis coefficients.
The resulting search dimension is independent of the prediction horizon.
Hence, a longer prediction horizon does not require more samples to maintain the sampling density.
For linear systems with quadratic costs, we derive a closed-loop stability condition in the infinite-sample limit. 
Under finite sampling, we provide a closed-form expression for the limit of the ratio of the effective sample size to the number of samples.
We also show that the upper bound of this limit for Laguerre-MPPI is no lower than that for conventional MPPI.
Numerical experiments demonstrate smoother inputs in obstacle avoidance and maintained performance over long horizons.
Moreover, they verify the lower bound on the number of samples given by the finite-sample analysis.
\end{abstract}

\section{INTRODUCTION}

Model Predictive Path Integral control (MPPI)~\cite{Williams2017,Williams2018} is a sampling-based Model Predictive Control (MPC) method.
It updates the input by a weighted average of randomly generated input sequences, with exponential weights determined by the cost of each input sequence.
MPPI has several advantages. 
Since MPPI requires no cost gradients, it directly handles nonlinear dynamics and non-convex or discontinuous costs.
Since the computations for individual samples are independent, MPPI is implemented efficiently on GPUs.
Furthermore, the input is updated using a weighted average of all samples, allowing information from the entire sample set to contribute to the update, unlike the random shooting method~\cite{Bakarac2018} and the cross-entropy method~\cite{Walker2026}.
Hence the computed input sequence tends to be smoother than those of these methods.
Owing to these advantages, MPPI has been applied to various control problems including whole-body control of legged robots~\cite{Padilla}, navigation of ground robots in cluttered and off-road environments~\cite{9834098,10606099}, agile flight and navigation of quadrotors~\cite{Minarik2024,Zhai2026}, aggressive driving on dirt tracks~\cite{7487277,Gandhi2021},
off-road driving over uneven terrain~\cite{10480553,Han-RSS-24}, and motion planning for autonomous driving~\cite{10601272}.

MPPI still has two open issues.
One is the degradation of control performance with longer prediction horizons.
Since MPPI samples input sequences over the prediction horizon, the search dimension grows linearly with the prediction horizon.
Hence, when limited computational resources prevent increasing the number of samples, extending the horizon makes the samples sparse in the search space due to the curse of dimensionality~\cite{bengtsson2008curse,agapiou2017importance} and degrades control performance.
In addition, when the number of samples is small relative to the search dimension, high-frequency components remain in the computed input and cause chattering and actuator wear~\cite{LP-MPPI}.
Existing studies on this issue fall broadly into three categories: methods that design the sampling distribution to increase the number of effective samples~\cite{Williams2018RSS,Gandhi2021,Trevisan2024,Sacks2022}, methods that modify how samples are drawn without changing the distribution~\cite{Mohamed2025,Kim2022}, and methods that reduce the search dimension itself~\cite{Padilla,Schramm2025}.

The other issue for MPPI is the lack of theoretical guarantees, in particular stability and performance analysis for the closed-loop system.
Existing research has focused on the evaluation of the sampling error~\cite{Yoon2022}, the convergence analysis of the optimization method~\cite{Yi2024}, and the increase in free energy caused by constraint violations or imperfect tracking performance~\cite{Gandhi2021}. The stability of the closed-loop system is regarded as an open problem~\cite{Honda2026}.
Yoon and Kim~\cite{Yoon2026} treated finite-sample MPPI for linear systems as a perturbation of LQR and showed practical exponential stability in expectation.
The number of samples required for their theory to hold remains unclear.

This paper develops a new control algorithm that reduces the search dimension of MPPI using Laguerre basis functions.
We then analyze the stability of the resulting closed-loop system and the sampling accuracy with a finite number of samples.
Laguerre basis functions form a discrete-time orthonormal basis and have been used in system identification~\cite{Wahlberg1991} and in MPC to reduce the number of optimization variables~\cite{Wang2001}. Motivated by their use in MPC,
we propose Laguerre-MPPI, which represents the sampled input of MPPI with Laguerre basis functions. Since Laguerre-MPPI samples the coefficients of the Laguerre basis instead of the input sequence, its search dimension is independent of the prediction horizon.
Hence, extending the prediction horizon does not require more samples to maintain the sampling density.
We derive a stability condition for the infinite-sample closed-loop system and show that closed-loop stability can be lost depending on the basis selection.
To evaluate the sampling error under finite samples, we present the closed form of the covariance of the sampling error and the limit of the ratio of the effective sample size (ESS) to the number of samples.
Based on these expressions, we provide a guideline on the number of samples required for the theory to hold, and show that the maximum of the asymptotic ESS ratio of Laguerre-MPPI over the initial estimate coefficient is no less than that of conventional MPPI.

The contributions of this paper are as follows.
\begin{enumerate}
\item We expand the input in Laguerre basis functions and develop Laguerre-MPPI (Section~\ref{sec:LMPPI}).
\item For linear systems with quadratic costs, we translate the closed-loop behavior into the behavior of a stochastic dynamical system to derive the stability condition in the infinite-sample limit (Section~\ref{sec:stability}).
\item We derive the asymptotic covariance of the sampling error and the asymptotic ESS ratio in closed form, and give a lower bound on the number of samples. We then show that, regardless of the basis selection, the maximum of the asymptotic ESS ratio of Laguerre-MPPI over the initial estimate coefficient is no less than that of conventional MPPI (Section~\ref{sec:ess}).
\end{enumerate}

\textit{Notation:}
The $m\times m$ identity matrix is denoted by $I_m$, the Kronecker product by $\otimes$, the vector obtained by stacking the columns of a matrix by $\mathrm{vec}(\cdot)$, the spectral radius by $\rho(\cdot)$, the Kronecker delta by $\delta_{ij}$, and the inverse $z$-transform by $\mathcal{Z}^{-1}$.
For a square matrix $X$, its determinant is denoted by $|X|$, and its positive definiteness (positive semidefiniteness) by $X\succ0$ ($X\succeq0$). For a vector $z$ and a matrix $Q$, the weighted squared norm is defined by $\|z\|_Q^2:=z^\top Q z$.
We denote by $\mathcal{N}(\mu,\Sigma)$ the Gaussian distribution with mean $\mu$ and covariance $\Sigma$, and by $\E_{\mathbb{F}}[\cdot]$ and $\mathrm{Cov}_{\mathbb{F}}[\cdot]$ the expectation and the covariance matrix with respect to a distribution $\mathbb{F}$. The probability density function of a distribution is denoted by the corresponding lowercase letter.
The Kullback--Leibler divergence is defined by $\DKL(\mathbb{F}\|\mathbb{H}):=\int f\ln (f/h)\,dX$.
Convergence in distribution is denoted by $\xrightarrow{d}$, and almost sure convergence by a.s.

\section{PRELIMINARIES: MPPI}\label{sec:MPPI}
Before presenting Laguerre-MPPI, we review the theory of MPPI on which it is built, following the information-theoretic derivation of Williams et al.~\cite{Williams2018}.
\subsection{Stochastic Optimal Control Problem}
Consider a general discrete-time system of the form
\begin{equation}\label{eq:sys}
x(k+1)=F(x(k),v(k)),\quad x(0)=x_0,
\end{equation}
where $x(k)\in\mathbb{R}^n$ is the state, $v(k)\in\mathbb{R}^m$ is the sampled input, and $F:\mathbb{R}^n\times\mathbb{R}^m\to\mathbb{R}^n$ represents the system dynamics.
The sampled input sequence up to the prediction horizon $T$ is denoted by
\begin{equation}\label{eq:V}
V:=[v(0)\,\cdots\,v(T-1)]\in\mathbb{R}^{m\times T}.
\end{equation}
The sampled input sequence $V$ is a random variable, and its columns $v(k)$ independently follow the Gaussian distribution $\mathcal{N}(u(k),\Sigma_v)$ with the mean input $u(k)$ and the covariance matrix $\Sigma_v\succ 0$.
The mean input sequence is denoted by
\begin{equation*}
U:=[u(0)\,\cdots\,u(T-1)]\in\mathbb{R}^{m\times T}.
\end{equation*}
Furthermore, we define the distribution of the sampled input sequence $V$ by $\mathbb{Q}_U:=\mathcal{N}(U,\Sigma_v)$, in which the columns $v(k)$ independently follow $\mathcal{N}(u(k),\Sigma_v)$, and call it the control distribution. The same notation is used for the other distributions of the sampled input sequence.

To derive the MPPI algorithm, we address the following stochastic optimal control problem.
\smallskip
\begin{Problem}[Stochastic Optimal Control]\label{prob:MPPI}
\begin{equation}\label{eq:SOC}
\min_{U\in \mathcal{U}}\ \E_{\mathbb{Q}_U}\Bigl[\phi(x(T))+\sum_{k=0}^{T-1}\mathcal{L}_k(x(k), u(k))\Bigr],
\end{equation}
where $\phi(x(T))$ is the terminal cost, $\mathcal{L}_k(x(k), u(k))$ is the stage cost of the state and input, and $\mathcal{U}$ is the feasible set of the mean input sequence $U$.
\end{Problem}
\smallskip
Since this problem is intractable due to the implicit form of $U$, we consider an approximation of it. We impose the following assumption.
\smallskip
\begin{assumption}\label{assumption:cost_decomposition}
The stage cost $\mathcal{L}_k(x(k), u(k))$ is
\begin{equation*}
  \mathcal{L}_k(x(k), u(k)) = c_k(x(k)) + b_k(u(k)),
\end{equation*}
where $c_k(x(k))$ is the state cost, and $b_k(u(k))$ is the input cost.
\end{assumption}
\smallskip

For brevity, we collect the state costs in \eqref{eq:SOC} as
\begin{equation}\label{eq:state_cost}
S(V):=\phi(x(T))+\sum_{k=0}^{T-1}c_k(x(k)).
\end{equation}
The state cost $S:\mathbb{R}^{m\times T}\to\mathbb{R}$ returns the cost of the trajectory generated from the sampled input sequence $V$.
Furthermore, we introduce the reference distribution $\tilde{\mathbb{Q}}:=\mathcal{N}(\tilde{U},\Sigma_v)$ with the reference input sequence $\tilde U$, to which the control distribution $\mathbb{Q}_U$ refers.
\subsection{Free Energy and Relaxed Stochastic Optimal Control Problem}
To solve \eqref{eq:SOC} approximately, we introduce the free energy~\cite{6426381}.
\smallskip
\begin{Definition}[Free energy]
  The following quantity is called the free energy of the control system composed of \eqref{eq:sys} and \eqref{eq:SOC}:
\begin{equation}\label{eq:free_energy}
    \mathcal{F}(S, \tilde{q},\lambda) := -\lambda\ln \left(\mathbb{E}_{\tilde{\mathbb{Q}}} \left[ \exp\left(-\frac{S(V)}{\lambda}\right) \right]\right),
\end{equation}
where $\lambda\in \mathbb{R}^+$ is called the temperature parameter.
\end{Definition}
\smallskip
Throughout this paper, we assume that the expectation in \eqref{eq:free_energy} is finite, which holds, e.g., when the state cost $S$ is bounded below.

We show that the free energy $\mathcal{F}$ is a lower bound of some cost function. To this end, we introduce an arbitrary probability distribution $\mathbb{Q}$ of the sampled input sequence $V$.
Then, the following lemma holds.
\smallskip
\begin{lemma}\label{lemma:free_energy_bound}
For any distribution $\mathbb{Q}$, the following inequality holds:
\begin{equation}\label{eq:free_energy_bound}
    \mathcal{F}(S, \tilde{q},\lambda) \leq \mathbb{E}_{\mathbb{Q}}\left[S(V)\right] + \lambda \DKL(\mathbb{Q}\|\tilde{\mathbb{Q}}).
\end{equation}
\end{lemma}
\smallskip
\begin{proof}
  See Appendix~\ref{app:free_energy_bound} (see also~\cite{Williams2018}).
\end{proof}
\smallskip
The first term on the right-hand side is the expected state cost, and the second term represents the discrepancy between the arbitrary distribution and the reference distribution through their KL divergence.
Intuitively, the first term is interpreted as a state cost and the second term as an input cost.
Hence, we let
\begin{equation}\label{eq:J(Q)}
    J(\mathbb{Q}) := \mathbb{E}_{\mathbb{Q}}\left[S(V)\right] + \lambda \DKL(\mathbb{Q}\|\tilde{\mathbb{Q}}).
\end{equation}
Note that the cost function $J(\mathbb{Q})$ is defined for an arbitrary distribution $\mathbb{Q}$, not only for Gaussian distributions.
To clarify the relation between the cost function $J(\mathbb{Q})$ and the objective function of Problem~\ref{prob:MPPI}, we first impose the following assumption.
\smallskip
\begin{assumption}\label{assumption:quadratic_input_cost}
The input cost $b_k(u(k))$ is a quadratic function of the mean input $u(k)$ given by
\begin{equation}\label{eq:input_cost}
    b_k(u(k)) = \frac{\lambda}{2} (u(k)-\tilde{u}(k))^\top\Sigma_{v}^{-1}(u(k)-\tilde{u}(k)).
\end{equation}
\end{assumption}
\smallskip

With this preparation, the following lemma shows that \eqref{eq:J(Q)} equals the cost function of \eqref{eq:SOC}.
\smallskip
\begin{lemma}\label{lemma:relaxed_SOC}
In the cost function $J(\mathbb{Q})$, let the reference distribution be $\tilde{\mathbb{Q}}=\mathcal{N}(\tilde U,\Sigma_v)$, and set the arbitrary distribution $\mathbb{Q}$ to the control distribution $\mathbb{Q}_U=\mathcal{N}(U,\Sigma_v)$. Under Assumptions~\ref{assumption:cost_decomposition} and \ref{assumption:quadratic_input_cost}, the following equation holds:
\begin{equation*}
    J(\mathbb{Q}_U) = \mathbb{E}_{\mathbb{Q}_U}\left[\phi(x(T))+\sum_{k=0}^{T-1} \mathcal{L}_k(x(k), u(k))\right].
\end{equation*}
\end{lemma}
\smallskip
\begin{proof}
  See Appendix~\ref{app:relaxed_SOC} (see also~\cite{Williams2018}).
  \end{proof}
  \smallskip

As stated in Lemma~\ref{lemma:relaxed_SOC}, Problem~\ref{prob:MPPI} is equivalently reduced to the following optimization problem over the set of control distributions $\mathcal{G}:=\{\mathcal{N}(U,\Sigma_v)\mid U\in\mathcal{U}\}$ as
\begin{equation}\label{eq:SOC_distribution}
  \min_{U\in\mathcal{U}} \mathbb{E}_{\mathbb{Q}_U}\left[\phi(x(T))+\sum_{k=0}^{T-1}\mathcal{L}_k(x(k),u(k))\right]
  = \min_{\mathbb{Q}\in\mathcal{G}} J(\mathbb{Q}).
\end{equation}
By further removing the constraint $\mathbb{Q}\in\mathcal{G}$, we formulate the following stochastic optimal control problem, which is a relaxation of \eqref{eq:SOC_distribution}:
\smallskip
\begin{Problem}[Relaxed Stochastic Optimal Control]\label{prob:relaxed_SOC}
\begin{equation*}
  \mathbb{Q}^* = \argmin_{\mathbb{Q}} J(\mathbb{Q})
\end{equation*}
\end{Problem}
\smallskip
Problem~\ref{prob:relaxed_SOC} is a variational problem of finding the distribution that attains the minimum cost without restricting the form of the distribution.
Since the feasible set expands from the set $\mathcal{G}$ of control distributions to all distributions, the following inequality holds between Problem~\ref{prob:relaxed_SOC} and the original Problem~\ref{prob:MPPI}:
\begin{equation*}
 \min_{\mathbb{Q}} J(\mathbb{Q}) \leq \min_{U\in\mathcal{U}} \mathbb{E}_{\mathbb{Q}_U}\left[\phi(x(T))+\sum_{k=0}^{T-1}\mathcal{L}_k(x(k),u(k))\right].
\end{equation*}

\subsection{Optimal Distribution\label{sec:optimal_distribution}}
This subsection gives the solution of Problem~\ref{prob:relaxed_SOC} through the following lemma.
\smallskip
\begin{lemma}\label{lemma:optimal_distribution}
The distribution $\mathbb{Q}^*$ with the probability density function
\begin{equation}\label{eq:optimal_control_distribution}
    q^*(V) := \frac{1}{\eta} \exp\left(-\frac{S(V)}{\lambda}\right) \tilde{q}(V)
\end{equation}
is the unique solution of Problem~\ref{prob:relaxed_SOC},
where $\eta$ is the normalization constant and is defined by
\begin{align}
    \eta &:= \int_{\mathbb{R}^{m\times T}} \exp\left(-\frac{S(V)}{\lambda}\right) \tilde{q}(V)\, dV\label{eq:eta}\\
         &= \mathbb{E}_{\tilde{\mathbb{Q}}} \left[ \exp\left(-\frac{S(V)}{\lambda}\right) \right].\notag
\end{align}
In addition, $J(\mathbb{Q}^*)=\mathcal{F}(S,\tilde{q},\lambda)$ holds.
\end{lemma}
\smallskip
\begin{proof}
  See Appendix~\ref{app:optimal_distribution} (see also~\cite{Williams2018}).
  \end{proof}
\smallskip

Hereafter, we call the distribution $\mathbb{Q}^*$, given by \eqref{eq:optimal_control_distribution}, the optimal distribution.
As \eqref{eq:optimal_control_distribution} shows, the optimal distribution $\mathbb{Q}^*$ has a complex shape even when the reference distribution is Gaussian, $\tilde{\mathbb{Q}}=\mathcal{N}(\tilde{U},\Sigma_v)$.
\subsection{Minimization of KL Divergence\label{sec:KL_minimization}}
A drawback of the optimal distribution $\mathbb{Q}^*$, characterized by \eqref{eq:optimal_control_distribution}, is that it is computationally intractable in general.
Moreover, $\mathbb{Q}^*$ is generally non-Gaussian and therefore cannot be recovered by restricting the search to the Gaussian control distributions $\mathbb{Q}_U = \mathcal{N}(U,\Sigma_v)$.  
Our goal is therefore to approximate $\mathbb{Q}^*$ by $\mathbb{Q}_U$, which is obtained by minimizing the KL divergence between $\mathbb{Q}_U$ and $\mathbb{Q}^*$.
\smallskip
\begin{Problem}[KL Divergence Minimization]\label{prob:KL_minimization}
\begin{equation}\label{eq:KL_minimization}
  U^* = \argmin_{U \in \mathcal{U}} \DKL(\mathbb{Q}^*\|\mathbb{Q}_U)
\end{equation}
\end{Problem}
\smallskip
When $\mathcal{U}=\mathbb{R}^{m\times T}$, Problem~\ref{prob:KL_minimization} is solvable in closed form, and the optimal input sequence $U^*$ in \eqref{eq:KL_minimization} is given by
\begin{equation}\label{eq:optimal_input_expectation}
    U^* = \mathbb{E}_{\mathbb{Q}^*}[V].
\end{equation}
The key implication of \eqref{eq:optimal_input_expectation} is that the optimal approximation to $\mathbb{Q}^*$ within the family of control distributions $\mathcal{N}(U,\Sigma_v)$ is determined solely by the mean of $\mathbb{Q}^*$.
Hence, the optimal input sequence $U^*$, the mean of the distribution $\mathbb{Q}_{U^*}$ that approximates $\mathbb{Q}^*$, is obtained by performing the integral:
\begin{equation}\label{eq:optimal_input_integral}
    U^* = \frac{1}{\eta}
  \int_{\mathbb{R}^{m\times T}} V \exp\left(-\frac{S(V)}{\lambda}\right)\tilde{q}(V)\,dV.
  \end{equation}

Note that the state cost $S(V)$, defined in \eqref{eq:state_cost}, is a complicated function and the integral is high-dimensional.
A promising approach to \eqref{eq:optimal_input_integral} is the use of sampling, that is, computing the expectation in \eqref{eq:optimal_input_expectation} by generating sampled input sequences that follow the optimal distribution $\mathbb{Q}^*$.
In the next subsection, we introduce an approach by importance sampling.
\subsection{Importance Sampling\label{sec:importance_sampling}}
Although no method is available for generating the sampled input sequence $V$ from $\mathbb{Q}^*$, the functional form $\exp\left(-\frac{S(V)}{\lambda}\right)\tilde{q}(V)$ is known. Hence importance sampling is applicable.
We therefore introduce a known initial estimate distribution $\hat{\mathbb{Q}}$ from which the sampled input sequence $V$ is readily generated.
The following theorem presents an alternative expression of $U^*$, given in \eqref{eq:optimal_input_expectation}.
\medskip
\begin{theorem}\label{theorem:MPPI_importance_sampling}
Let the initial estimate distribution and the reference distribution be given by $\hat{\mathbb{Q}}=\mathcal{N}(\hat U,\Sigma_v)$ and $\tilde{\mathbb{Q}}=\mathcal{N}(\tilde U,\Sigma_v)$, respectively. Then, the optimal input sequence $U^*$, given in \eqref{eq:optimal_input_expectation}, is expressed with the weight $w(V)$ as
\begin{align}
  U^* &= \frac{\mathbb{E}_{\hat{\mathbb{Q}}}[w\,V]}{\mathbb{E}_{\hat{\mathbb{Q}}}[w]},\label{eq:Ustar}\\
  w(V) &:= \exp\left(-\frac{S(V)}{\lambda}-\sum_{k=0}^{T-1}(\hat u(k)-\tilde u(k))^\top\Sigma_v^{-1}v(k)\right).\label{eq:w}
\end{align}
\end{theorem}
\medskip
\begin{proof}
  See Appendix~\ref{app:MPPI_importance_sampling} (see also~\cite{Williams2018}).
\end{proof}
\smallskip
Since $\hat{\mathbb{Q}}$ is known and the weight $w$ is computable, we evaluate \eqref{eq:Ustar} by drawing samples from $\hat{\mathbb{Q}}$ and thereby compute the optimal input sequence $U^*$ in \eqref{eq:optimal_input_expectation} indirectly.

To summarize this section, the original stochastic control problem, stated as Problem~\ref{prob:MPPI}, is approximated by Problem~\ref{prob:KL_minimization}.
MPPI provides an efficient method for solving Problem~\ref{prob:KL_minimization} by computing a sample-based approximation of the solution characterized by Theorem~\ref{theorem:MPPI_importance_sampling}.
The detailed procedure is summarized in Algorithm~\ref{algorithm:mppi}. The input sequence $U^\dagger$ obtained using a finite number of samples approximates the optimal solution $U^*$ of Problem~\ref{prob:KL_minimization}, given in \eqref{eq:Ustar}.
\begin{algorithm}[t]
\caption{MPPI}
\label{algorithm:mppi}
\begin{algorithmic}[1]
\Require $M$, $T$, $\lambda$, $\Sigma_v$, $\tilde U$, $\hat U$
\For{$k=0,1,2,\ldots$}
  \State Generate $\{V^i\}_{i=1}^M$ from $\hat{\mathbb{Q}}=\mathcal{N}(\hat U,\Sigma_v)$
  \For{$i=1,\ldots,M$}
    \State Apply $V^i$ to \eqref{eq:sys} and compute $S(V^i)$
    \State Compute $w(V^i)$ by \eqref{eq:w}
  \EndFor
  \State $U^\dagger\gets\frac{\sum_{i=1}^M w(V^i)V^i}{\sum_{i=1}^M w(V^i)}$
  \State Apply the first column $u^\dagger(0)$ of $U^\dagger$ to the system
  \State Update $\hat U$
\EndFor
\end{algorithmic}
\end{algorithm}

Finally, we present a remark that motivates the main results developed in the subsequent sections.
The search space of MPPI is $mT$-dimensional, which grows linearly with the prediction horizon $T$.
For a fixed number of samples $M$, extending the horizon makes the samples increasingly sparse in this space, leading to degraded control performance.
To address this issue, Section~\ref{sec:LMPPI} proposes a control method whose search dimension is independent of $T$.
Section~\ref{sec:ess} quantifies this degradation using the ESS.

\section{LAGUERRE-MPPI AND ALGORITHM}\label{sec:LMPPI}
To mitigate the performance degradation of MPPI over long prediction horizons, we expand the control input in Laguerre basis functions.
Instead of drawing the sampled input sequence $V$ directly, we generate it from coefficients drawn in the coefficient space, whose dimension is independent of the prediction horizon $T$.
Section~\ref{sec:basis} introduces the Laguerre basis functions, and Section~\ref{sec:expansion} expands the input in the basis and derives the linear relationship between the coefficients and the inputs.
Section~\ref{sec:LMPPI_def} formulates the stochastic optimal control problem in the coefficient space, reconstructs the theory of MPPI in that space, and gives the algorithm of Laguerre-MPPI.
Section~\ref{sec:coef_cov} shows that the algorithm of Laguerre-MPPI reduces to that of MPPI when the covariance matrix in the coefficient space satisfies a specific condition.
Section~\ref{sec:related} compares Laguerre-MPPI with related studies.

\subsection{Discrete-Time Laguerre Basis Functions}\label{sec:basis}
\begin{figure}[t]
  \centering
  \includegraphics[width=\linewidth]{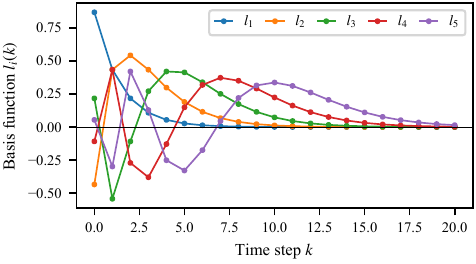}
  \caption{Example of discrete-time Laguerre basis functions with the pole $a=0.5$.}
  \label{fig:Laguerre_basis}
\end{figure}

The discrete-time Laguerre basis functions with pole $a\in[0,1)$ are defined by the $z$-transform~\cite{broome1965discrete,Wahlberg1991}
\begin{equation}\label{eq:Lz}
\Gamma_i(z):=\frac{\sqrt{1-a^2}}{z-a}\left(\frac{1-az}{z-a}\right)^{i-1},
\end{equation}
where $i\in \{1,2,\ldots\}$ is the order of the basis function.
Since the all-pass factor has unit magnitude on the unit circle, all basis functions share the magnitude response $|\Gamma_i(e^{\mathrm{j}\omega})|=\sqrt{1-a^2}/|e^{\mathrm{j}\omega}-a|$, which has low-pass characteristics for any pole $a>0$.
We define the basis functions in the time domain as
\begin{equation}\label{eq:li}
l_i(k):=\mathcal{Z}^{-1}[z\Gamma_i(z)].
\end{equation}
Parseval's theorem and the all-pass property of the factor $(1-az)/(z-a)$ yield $\sum_{k=0}^{\infty}l_i(k)l_j(k)=\delta_{ij}$~\cite{Wahlberg1991}.
We stack $N$ basis functions as
\begin{equation*}
L(k):=[l_1(k)\,\cdots\,l_N(k)]^\top
\end{equation*}
and let
\begin{equation*}
\beta:=1-a^2.
\end{equation*}
Then, $L(k)$ evolves as follows~\cite{Wang2001}:
\begin{equation}\label{eq:Lrec}
L(k+1)=\Omega L(k),
\end{equation}
where $\Omega$ is given by
\begin{equation}\label{eq:Omega}
\Omega:=\begin{bmatrix}
a&0&\cdots&0\\
\beta&a&\cdots&0\\
\vdots&\ddots&\ddots&\vdots\\
(-a)^{N-2}\beta&\cdots&\beta&a
\end{bmatrix}
\end{equation}
and $L(0)$ is given by
\begin{equation*}
L(0):=\sqrt{\beta}\,[1\ \ {-a}\ \ a^2\ \cdots\ (-a)^{N-1}]^\top.
\end{equation*}
To illustrate the Laguerre basis, \figref{fig:Laguerre_basis} shows the basis functions up to the order $N=5$ for the pole $a=0.5$. As the figure illustrates, the Laguerre basis has the following two properties.
\begin{itemize}
  \item Since each basis function decays exponentially in time, the basis represents the inputs in the first part of the prediction horizon finely and those in the latter part coarsely.
  \item The peak of a basis function is delayed as its order increases.
\end{itemize}
The first property originates from the first-order low-pass factor $\sqrt{1-a^2}/(z-a)$ in \eqref{eq:Lz}, which is common to all basis functions.
This factor alone yields the first basis function $l_1(k)=\sqrt{\beta}\,a^k$, which decays exponentially with ratio $a$, and the higher-order basis functions inherit this decay.
The second property originates from the all-pass factor $(1-az)/(z-a)$, which is multiplied once more for each increase in the order~\cite{VANDENHOF2000445}.
Since the all-pass factor does not change the magnitude response, it only adds phase lag. At low frequencies, this lag corresponds to a time delay of $(1+a)/(1-a)$ steps.
Hence each increase in the order shifts the peak of the basis function later in time.
As a consequence of the second property, the low-order basis functions concentrate on the beginning of the prediction horizon, 
and a small number of low-order basis functions suffices to represent the inputs at the beginning of the prediction horizon.

\subsection{Laguerre Basis Expansion of Control Inputs}\label{sec:expansion}
Using the Laguerre basis introduced above, we expand the inputs in the basis and derive the linear relationship between the coefficients and the inputs.
We focus on the $r$-th channel of $v(k)$, denoted by $v_r(k)$, and present its expansion:
\begin{equation}\label{eq:sel1d}
v_r(k)=\sum_{i=1}^{N}s_{ir}\theta_{ir}l_i(k),
\end{equation}
where $\theta_{ir}\in \mathbb{R}$ is the coefficient, and $s_{ir} \in \{0,1\}$ is the indicator.
If $s_{ir}=1$, the $i$-th basis function is used for the expansion, and if $s_{ir}=0$, it is omitted.
Suppose that $d$ basis functions are selected from $N$ candidates $l_1(k),\ldots,l_N(k)$, i.e., it holds that
\begin{equation*}
\sum_{i=1}^{N}s_{ir}=d.
\end{equation*}
Let $\mathcal{I}_r:=\{i_{1r},\ldots,i_{dr}\}$ be the index set of the basis functions selected in the $r$-th channel.
Furthermore, letting $e_i\in\mathbb{R}^N$ be the $i$-th unit vector, we define
\begin{equation*}
\Sel_r:=[e_{i_{1r}}\,\cdots\,e_{i_{dr}}]\in\{0,1\}^{N\times d}. 
\end{equation*}
Then, we rewrite the expansion \eqref{eq:sel1d} in matrix form as
\begin{equation*}
v_r(k)=L(k)^\top\Sel_r\Sel_r^\top \theta_r,
\end{equation*}
where
\begin{equation*}
\theta_r:=[\theta_{1r}\,\cdots\,\theta_{Nr}]^\top.
\end{equation*}
Furthermore, letting
\begin{align*}
\vartheta&:=[\theta_1^\top\,\cdots\,\theta_m^\top]^\top\in\mathbb{R}^{mN},\\
\Sel&:=\mathrm{diag}(\Sel_1,\ldots,\Sel_m)\in \{0,1\}^{mN\times md},\\
\vartheta_{\Sel}&:=\Sel^\top\vartheta\in\mathbb{R}^{md},
\end{align*}
we have the expression of the sampled input $v(k)$ as
\begin{equation}\label{eq:varthetaS}
v(k)=P_{\Sel}(k)\vartheta_{\Sel},
\end{equation}
where $P_{\Sel}(k)$ is given by
\begin{equation}\label{eq:vexp}
P_{\Sel}(k):=(I_m \otimes L(k)^\top)\Sel \in\mathbb{R}^{m\times md}.
\end{equation}
For simplicity of notation, we present the expansion of the sampled input sequence $V$, given in \eqref{eq:V}, in vector form. We define the sampled input vector by
\begin{equation*}
\mathcal{V}:=\mathrm{vec}(V^\top).
\end{equation*}
Then, we have the expression
\begin{equation*}
\mathcal{V} =\mathcal{Q}_\Sel\vartheta_{\Sel},
\end{equation*}
where
\begin{align}
\mathcal{Q}_\Sel&:=(I_m \otimes \Phi^\top)\Sel\in\mathbb{R}^{mT\times md},\label{eq:QSel}\\
\Phi&:=[L(0)\,\cdots\,L(T-1)].\label{eq:Phi}
\end{align}

In a similar manner to $v(k)$, we present the expansions of the mean input $u(k)$, the reference input $\tilde{u}(k)$, and the initial estimate input $\hat{u}(k)$ as
\begin{align}
u(k)&=P_{\Sel}(k)\varpi_{\Sel},\label{eq:u}\\
\tilde u(k)&=P_{\Sel}(k)\tilde\varpi_{\Sel},\label{eq:ut}\\
\hat u(k)&=P_{\Sel}(k)\hat\varpi_{\Sel},\label{eq:uh}
\end{align}
where the coefficients $\varpi_{\Sel},\tilde\varpi_{\Sel},\hat\varpi_{\Sel}\in\mathbb{R}^{md}$ are called the mean coefficient, the reference coefficient, and the initial estimate coefficient, respectively.

We clarify the properties of $\Phi$, defined in \eqref{eq:Phi}, and discuss the relation between the sampled input vector $\mathcal{V}$ and the selected coefficient $\vartheta_{\Sel}$. To this end, we impose the following assumption regarding the prediction horizon $T$ and the number of candidate basis functions $N$.
\smallskip
\begin{assumption}\label{assumption:TN}
  $T\ge N$.
\end{assumption}
\smallskip
Under this assumption, the following lemma holds.
\smallskip
\begin{lemma}\label{lem:Phi}
Under Assumption~\ref{assumption:TN}, the following three statements hold.
\begin{enumerate}
\item The pair $(\Omega,L(0))$ is controllable.
\item $\mathrm{rank}(\Phi)=N$.
\item $\Phi\Phi^\top\succ 0$.
\end{enumerate}
\end{lemma}
\smallskip
\begin{proof}
 \textit{1)}
We show that the controllability matrix
\begin{equation*}
\mathcal{M}_{\Omega}:=[L(0)\ \Omega L(0)\ \cdots\ \Omega^{N-1}L(0)]\in\mathbb{R}^{N\times N}
\end{equation*}
is invertible. Using the recursion \eqref{eq:Lrec} repeatedly, we have $L(k)=\Omega^kL(0)$.
The matrix $\Omega$, defined in \eqref{eq:Omega}, is decomposed as $\Omega=aI_N+D$ with a strictly lower triangular matrix $D$.
The $(i,j)$ entry of the matrix $D$ is $D_{ij}=\beta(-a)^{i-j-1}$ for $i>j$. In particular, $D_{i,i-1}=\beta>0$ for $i=j+1$.
Since the matrices $aI_N$ and $D$ commute, the binomial theorem gives
\begin{equation*}
L(k)=\Omega^kL(0)=\sum_{i=0}^{k}\binom{k}{i}a^{k-i}D^i L(0).
\end{equation*}
Therefore, we have the expression
\begin{equation*}
\mathcal{M}_{\Omega}=\mathcal{M}_{D}\mathcal{A},
\end{equation*}
where the matrices $\mathcal{M}_{D}\in \mathbb{R}^{N\times N}$ and $\mathcal{A} \in\mathbb{R}^{N\times N}$ are given by
\begin{align*}
\mathcal{M}_{D}&:=
\begin{bmatrix}
L(0)&DL(0)&\cdots&D^{N-1}L(0)
\end{bmatrix},\\
\mathcal{A}&:=\begin{bmatrix}
1&a&\cdots&a^{N-1}\\
0&1&\cdots&\binom{N-1}{1}a^{N-2}\\
\vdots&\vdots&\ddots&\vdots\\
0&0&\cdots&1
\end{bmatrix},
\end{align*}
respectively.
Since the matrix $\mathcal{A}$ is upper triangular and its diagonal entries are all 1, it is invertible.
Hence the controllability matrix $\mathcal{M}_{\Omega}$ is invertible if and only if the matrix $\mathcal{M}_{D}$ is invertible.
Since the matrix $D^k$ is strictly lower triangular, the first $k$ entries of the vector $D^kL(0)$ are zero, and
\begin{equation*}
D^kL(0)=[0\ \cdots\ 0\ \beta^{k+\frac{1}{2}}\ *\ \cdots\ *]^\top.
\end{equation*}
Therefore the matrix $\mathcal{M}_{D}$ is lower triangular, and its diagonal entries $\beta^{k+\frac{1}{2}}, k \in \{0,\ldots,N-1\}$, are positive since $\beta>0$ for any pole $a\in[0,1)$.
Hence the matrix $\mathcal{M}_{D}$ is invertible.
Consequently, the pair $(\Omega,L(0))$ is controllable.

\textit{2)}
From Assumption~\ref{assumption:TN}, the first $N$ columns of $\Phi=[L(0)\ \cdots\ L(T-1)]$, defined in \eqref{eq:Phi}, form the controllability matrix $\mathcal{M}_{\Omega}$.
Since the controllability matrix $\mathcal{M}_{\Omega}$ is invertible as shown in 1), $\mathrm{rank}(\Phi)\ge N$.
Thus, $\mathrm{rank}(\Phi)=N$.

\textit{3)}
It follows from 2) that $\Phi\Phi^\top\succ 0$ holds.
\end{proof}
\smallskip
Using Lemma~\ref{lem:Phi}, we derive the following lemma.
\smallskip
\begin{lemma}\label{lem:P_sel}
Under Assumption~\ref{assumption:TN}, $\mathcal{Q}_{\Sel}$, defined in \eqref{eq:QSel}, is of full column rank.
\end{lemma}
\smallskip
\begin{proof}
By 2) of Lemma~\ref{lem:Phi}, $I_m\otimes\Phi^\top$ is of full column rank. Since the columns of the selection matrix $\Sel$ are distinct unit vectors, $\mathcal{Q}_{\Sel}=(I_m\otimes\Phi^\top)\Sel$ is also of full column rank.
\end{proof}
\smallskip

As stated in Lemma~\ref{lem:P_sel}, $\mathcal{Q}_{\Sel}$ is of full column rank, and hence the selected coefficient $\vartheta_{\Sel}$ and the sampled input vector $\mathcal{V}$ are in one-to-one correspondence.

The number of selected basis functions typically satisfies $d<T$. Then, the sampled input vector $\mathcal{V}$ takes values in the $md$-dimensional subspace $\operatorname{Im}(\mathcal{Q}_{\Sel})$ of $\mathbb{R}^{mT}$ and therefore has no probability density on $\mathbb{R}^{mT}$.
Hence the derivation of MPPI in Section~\ref{sec:MPPI}, which assumes that the sampled input sequence has a probability density on $\mathbb{R}^{mT}$, is not applicable to the sampled input sequence expanded in the Laguerre basis.
We therefore search for a solution to the control problem indirectly in the coefficient space of $\vartheta_{\Sel}$, rather than directly in the input space of $\mathcal{V}$.

\subsection{Laguerre-MPPI}\label{sec:LMPPI_def}
\subsubsection{Optimal control problem in the coefficient space}\label{sec:coef_SOC}
Recall that MPPI solves Problem~\ref{prob:MPPI} by optimizing the mean $U$ of the distribution of the sampled input sequence $V$.
Instead, in this subsection, we optimize the mean $\varpi_{\Sel}$ of the distribution of the selected coefficient $\vartheta_{\Sel}$.
To this end, we first formulate a counterpart of Problem~\ref{prob:MPPI} in the coefficient space.

Let $\Sigma_{\Sel}\in\mathbb{R}^{md\times md}$ with $\Sigma_{\Sel}\succ0$ be the covariance matrix of the selected coefficient.
Analogously to the control distribution $\mathbb{Q}_U$, the reference distribution $\tilde{\mathbb{Q}}$, and the initial estimate distribution $\hat{\mathbb{Q}}$, introduced in Section~\ref{sec:MPPI},
we define the corresponding distributions in the coefficient space as
\begin{align*}
\mathbb{Q}_{\varpi_{\Sel}}&:=\mathcal{N}(\varpi_{\Sel},\Sigma_{\Sel}),\\
\tilde{\mathbb{Q}}_{\Sel}&:=\mathcal{N}(\tilde\varpi_{\Sel},\Sigma_{\Sel}),\\
\hat{\mathbb{Q}}_{\Sel}&:=\mathcal{N}(\hat\varpi_{\Sel},\Sigma_{\Sel}),
\end{align*}
where the mean coefficient $\varpi_{\Sel}$, the reference coefficient $\tilde\varpi_{\Sel}$, and the initial estimate coefficient $\hat\varpi_{\Sel}$ are those in \eqref{eq:u}--\eqref{eq:uh}.
In the coefficient space, we call $\mathbb{Q}_{\varpi_{\Sel}}$, $\tilde{\mathbb{Q}}_{\Sel}$, and $\hat{\mathbb{Q}}_{\Sel}$ the control distribution, the reference distribution, and the initial estimate distribution, respectively.
Since $\Sigma_{\Sel}$ is positive definite, these distributions have probability densities on $\mathbb{R}^{md}$. In addition,
let the function $S_{\vartheta_{\Sel}}:\mathbb{R}^{md}\to\mathbb{R}$ be given by
\begin{equation}\label{eq:Stheta}
S_{\vartheta_{\Sel}}(\vartheta_{\Sel}):=S(V(\vartheta_{\Sel})),
\end{equation}
which returns the cost of the sampled input sequence $V$ generated from the selected coefficient $\vartheta_{\Sel}$ by \eqref{eq:varthetaS}.
Furthermore, we define the coefficient cost $b_{\Sel}(\varpi_{\Sel})$ instead of the input cost $b_k(u(k))$, given in \eqref{eq:input_cost}, as
\begin{equation}\label{eq:input_cost_coef}
b_{\Sel}(\varpi_{\Sel}):=\frac{\lambda}{2}\|\varpi_{\Sel}-\tilde\varpi_{\Sel}\|_{\Sigma_{\Sel}^{-1}}^2.
\end{equation}
With this preparation, we formulate the stochastic optimal control problem in the coefficient space as follows.
\smallskip
\begin{Problem}[Coefficient-Space Optimal Control]\label{prob:coef_SOC}
\begin{equation}\label{eq:SOC_coef}
\min_{\varpi_{\Sel}\in\mathbb{R}^{md}}\ \E_{\mathbb{Q}_{\varpi_{\Sel}}}\Bigl[\phi(x(T))+\sum_{k=0}^{T-1}c_k(x(k))\Bigr]+b_{\Sel}(\varpi_{\Sel}).
\end{equation}
\end{Problem}
\smallskip

The first term of the cost function is the expected state cost of the sampled input sequence $V$ generated by \eqref{eq:varthetaS} from the selected coefficient $\vartheta_{\Sel}$ that follows the control distribution $\mathbb{Q}_{\varpi_{\Sel}}$.
The second term penalizes the discrepancy between the mean coefficient $\varpi_{\Sel}$ and the reference coefficient $\tilde{\varpi}_{\Sel}$.

\subsubsection{MPPI in the coefficient space}\label{sec:coef_MPPI}
We solve Problem~\ref{prob:coef_SOC} approximately by a sampling-based approach.
In a similar manner to the derivation of MPPI, we derive the following key theorem on the approximation of the solution to Problem~\ref{prob:coef_SOC}.
For a distribution $\mathbb{Q}_{\Sel}$ with a probability density on $\mathbb{R}^{md}$, let the cost function in the coefficient space be
\begin{equation*}
J(\mathbb{Q}_{\Sel}):=\E_{\mathbb{Q}_{\Sel}}\bigl[S_{\vartheta_{\Sel}}(\vartheta_{\Sel})\bigr]+\lambda\DKL(\mathbb{Q}_{\Sel}\|\tilde{\mathbb{Q}}_{\Sel}).
\end{equation*}
Then, consider the problem of $\min_{\mathbb{Q}_{\Sel}}J(\mathbb{Q}_{\Sel})$, which is a relaxation of Problem~\ref{prob:coef_SOC}.
For this relaxed problem, we have the following theorem.
\medskip
\begin{theorem}\label{thm:coefMPPI}
The following four statements hold.
\begin{enumerate}
\item The cost function $J(\mathbb{Q}_{\varpi_{\Sel}})$ equals the objective function of \eqref{eq:SOC_coef}.
\item The unique solution to the relaxed problem $\min_{\mathbb{Q}_{\Sel}}J(\mathbb{Q}_{\Sel})$ is given by the optimal distribution $\mathbb{Q}_{\Sel}^*$ with the probability density function
\begin{equation}\label{eq:qstar}
q_{\Sel}^*(\vartheta_{\Sel}):=\frac{1}{\eta_{\Sel}}\exp\Bigl(-\frac{S_{\vartheta_{\Sel}}(\vartheta_{\Sel})}{\lambda}\Bigr)\tilde q_{\Sel}(\vartheta_{\Sel}),
\end{equation}
where $\eta_{\Sel}$ is the normalization constant and is given by
\begin{equation*}
\eta_{\Sel}:=\E_{\tilde{\mathbb{Q}}_{\Sel}}[\exp(-S_{\vartheta_{\Sel}}(\vartheta_{\Sel})/\lambda)].
\end{equation*}
In addition, it holds that
\begin{equation*}
J(\mathbb{Q}_{\Sel}^*)=-\lambda\ln\eta_{\Sel}=:\mathcal{F}_{\Sel},
\end{equation*}
which equals the free energy in the coefficient space.
\item The solution to the KL divergence minimization problem $\min_{\varpi_{\Sel}}\DKL(\mathbb{Q}_{\Sel}^*\|\mathbb{Q}_{\varpi_{\Sel}})$ is given in closed form by
\begin{equation*}
\varpi_{\Sel}^*=\E_{\mathbb{Q}_{\Sel}^*}[\vartheta_{\Sel}].
\end{equation*}
\item For the initial estimate distribution $\hat{\mathbb{Q}}_{\Sel}$, the optimal coefficient $\varpi_{\Sel}^*$ is expressed with the weight $w_{\Sel}(\vartheta_{\Sel})$ as
\begin{align}
\varpi_{\Sel}^*&=\frac{\E_{\hat{\mathbb{Q}}_{\Sel}}[w_{\Sel}\vartheta_{\Sel}]}{\E_{\hat{\mathbb{Q}}_{\Sel}}[w_{\Sel}]},\label{eq:LIS}\\
w_{\Sel}(\vartheta_{\Sel})&:=\exp\Bigl(-\frac{S_{\vartheta_{\Sel}}(\vartheta_{\Sel})}{\lambda}-(\hat\varpi_{\Sel}-\tilde\varpi_{\Sel})^\top\Sigma_{\Sel}^{-1}\vartheta_{\Sel}\Bigr).\label{eq:Lw}
\end{align}
\end{enumerate}
\end{theorem}
\medskip
\begin{proof}
Each statement follows from the derivation of its counterpart in Section~\ref{sec:MPPI} with the sampled input sequence $V$, the mean input sequence $U$,
and the integration domain $\mathbb{R}^{m\times T}$ replaced by the selected coefficient $\vartheta_{\Sel}$, the mean coefficient $\varpi_{\Sel}$, and $\mathbb{R}^{md}$, respectively.
The derivations in Section~\ref{sec:MPPI} use only two properties of the distributions. One is that the distributions have probability densities,
and the other is that the control distribution $\mathbb{Q}_{U}$, the reference distribution $\tilde{\mathbb{Q}}$, and the initial estimate distribution $\hat{\mathbb{Q}}$ are Gaussian with a common covariance matrix.
Both properties hold in the coefficient space, since the covariance matrix $\Sigma_{\Sel}$ is positive definite and is shared by the distributions $\mathbb{Q}_{\varpi_{\Sel}}$, $\tilde{\mathbb{Q}}_{\Sel}$, and $\hat{\mathbb{Q}}_{\Sel}$.
The only difference in the computation is that the sum $\sum_{k=0}^{T-1}(\cdot)^\top\Sigma_v^{-1}(\cdot)$ of quadratic forms over time becomes the single quadratic form $(\cdot)^\top\Sigma_{\Sel}^{-1}(\cdot)$.

Statement 1) follows from the proof of Lemma~\ref{lemma:relaxed_SOC} with Assumption~\ref{assumption:quadratic_input_cost} replaced by the definition \eqref{eq:input_cost_coef}.
The derivation gives $\lambda\DKL(\mathbb{Q}_{\varpi_{\Sel}}\|\tilde{\mathbb{Q}}_{\Sel})=\frac{\lambda}{2}\|\varpi_{\Sel}-\tilde\varpi_{\Sel}\|_{\Sigma_{\Sel}^{-1}}^2$, which equals the coefficient cost $b_{\Sel}(\varpi_{\Sel})$.

Statement 2) follows from the proof of Lemma~\ref{lemma:optimal_distribution}.

Statement 3) follows since the KL divergence $\DKL(\mathbb{Q}_{\Sel}^*\|\mathbb{Q}_{\varpi_{\Sel}})$ equals $\frac12\E_{\mathbb{Q}_{\Sel}^*}[\|\vartheta_{\Sel}-\varpi_{\Sel}\|_{\Sigma_{\Sel}^{-1}}^2]$ up to a constant independent of $\varpi_{\Sel}$, and its minimizer is the mean $\E_{\mathbb{Q}_{\Sel}^*}[\vartheta_{\Sel}]$ of the optimal distribution.

Statement 4) follows from the proof of Theorem~\ref{theorem:MPPI_importance_sampling}, which gives the density ratio
\begin{equation*}
\ln\frac{\tilde q_{\Sel}(\vartheta_{\Sel})}{\hat q_{\Sel}(\vartheta_{\Sel})}=-(\hat\varpi_{\Sel}-\tilde\varpi_{\Sel})^\top\Sigma_{\Sel}^{-1}\vartheta_{\Sel}+\mathrm{const.},
\end{equation*}
where the constant term does not depend on $\vartheta_{\Sel}$.
Therefore, \eqref{eq:LIS} and \eqref{eq:Lw} follow.
\end{proof}
\smallskip

Statement 4) of Theorem~\ref{thm:coefMPPI} enables us to develop a new MPPI scheme that samples in the coefficient space.
Before presenting its details, we compare it with the conventional MPPI scheme given in Algorithm~\ref{algorithm:mppi} as follows.
\begin{enumerate}[label=\alph*)]
\item \textit{Decision variables}: Problem~\ref{prob:MPPI} optimizes the mean input sequence $U$ directly, whereas Problem~\ref{prob:coef_SOC} optimizes the mean coefficient $\varpi_{\Sel}$, which determines $U$ through \eqref{eq:u}. This parameterization restricts $U$ to the $md$-dimensional subspace $\operatorname{Im}(\mathcal{Q}_{\Sel})$.
\item \textit{Random variables}: The random variable of MPPI is the sampled input sequence $V\in\mathbb{R}^{m\times T}$, whereas that of the coefficient-space formulation is the selected coefficient $\vartheta_{\Sel}\in\mathbb{R}^{md}$.
Both formulations evaluate the state cost with the same function $S$, defined in \eqref{eq:state_cost}, and the coefficient-space formulation evaluates it through $V$ generated from $\vartheta_{\Sel}$.
\item \textit{Input costs}: MPPI uses the input cost \eqref{eq:input_cost}, whereas the coefficient-space formulation uses the coefficient cost \eqref{eq:input_cost_coef}.
\item \textit{Covariance}: MPPI assigns the covariance matrix $\Sigma_v$ to the sampled input $v(k)$ independently at each time, whereas the covariance matrix $\Sigma_{\Sel}$ of the selected coefficient is an arbitrary positive definite matrix and may include correlations across basis functions and channels.
\end{enumerate}

\subsubsection{Laguerre-MPPI algorithm}\label{sec:LMPPI_alg}
\begin{figure*}[t]
\centering
\includegraphics[width=0.7\linewidth]{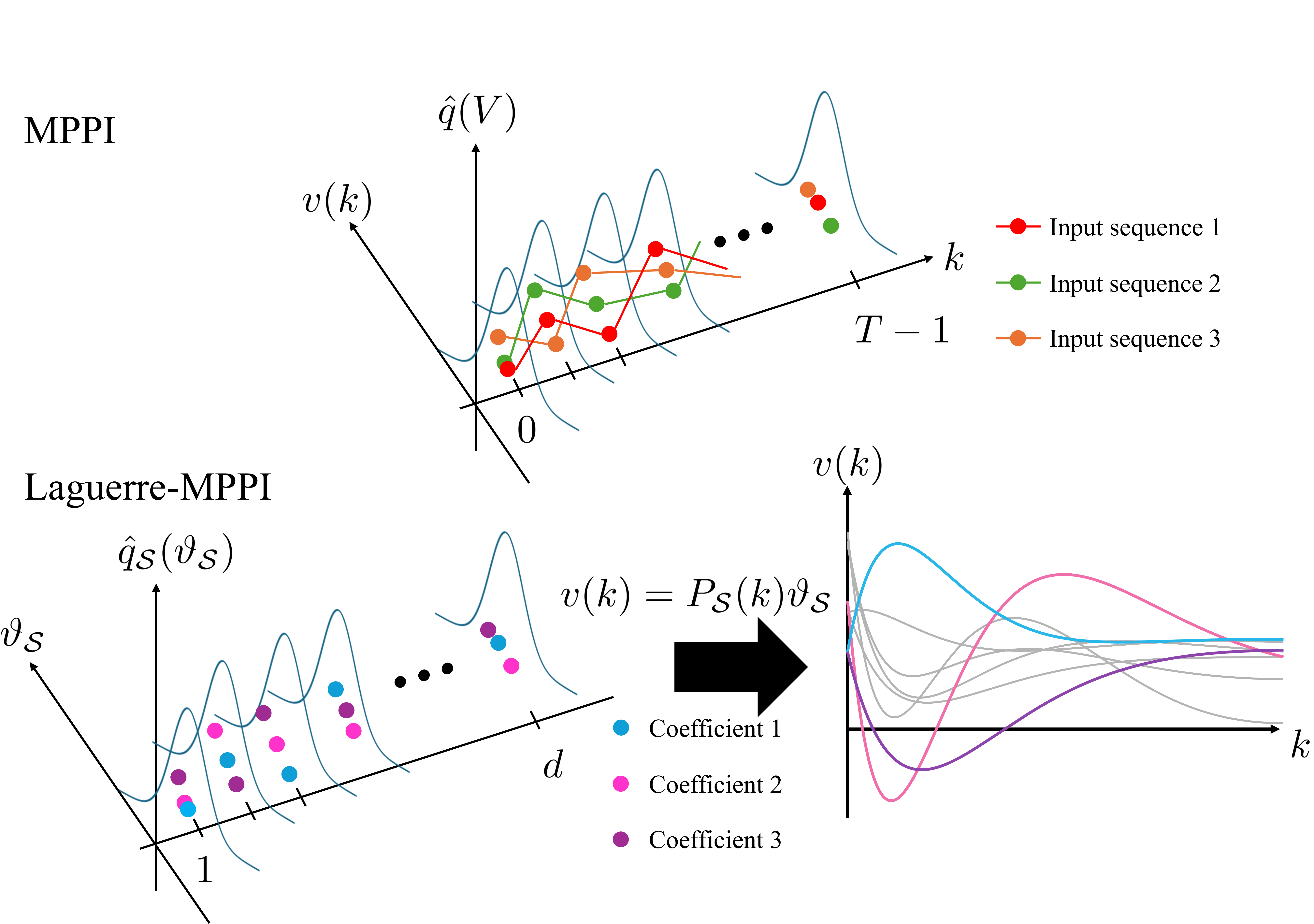}
\caption{Comparison between MPPI and Laguerre-MPPI.}
\label{fig:LMPI}
\end{figure*}

We now define Laguerre-MPPI as the sampling-based control method that approximately solves Problem~\ref{prob:coef_SOC} based on Theorem~\ref{thm:coefMPPI}.
The detailed procedure is summarized in Algorithm~\ref{algorithm:lmppi}.
\figref{fig:LMPI} illustrates the difference in the generation of the sampled input sequence between Laguerre-MPPI and conventional MPPI.
In conventional MPPI, the sampled input $v(k)$ at each time independently follows $\mathcal{N}(\hat u(k),\Sigma_v)$. Hence the sampled input sequence fluctuates from one time step to the next and is not smooth.
In Laguerre-MPPI, the selected coefficient is sampled instead, and each sampled input sequence is a linear combination of the selected basis functions $l_i(k)$, defined in \eqref{eq:li}.
Since each basis function has low-pass characteristics for any pole $a>0$, both the sampled input sequence and the updated mean input sequence are smooth in time.
The dimension of the selected coefficient is $md$, which does not depend on the prediction horizon $T$.
\begin{algorithm}[t]
\caption{Laguerre-MPPI}
\label{algorithm:lmppi}
\begin{algorithmic}[1]
\Require $M$, $T$, $\lambda$, $a$, $\Sel$, $\Sigma_{\Sel}$, $\tilde\varpi_{\Sel}$, $\hat\varpi_{\Sel}$
\For{$k=0,1,2,\ldots$}
  \State Generate $\{\vartheta_{\Sel}^i\}_{i=1}^M$ from $\hat{\mathbb{Q}}_{\Sel}=\mathcal{N}(\hat\varpi_{\Sel},\Sigma_{\Sel})$
  \For{$i=1,\ldots,M$}
    \State Convert $\vartheta_{\Sel}^i$ into $V^i$ by \eqref{eq:varthetaS}
    \State Apply $V^i$ to \eqref{eq:sys} and compute $S_{\vartheta_{\Sel}}(\vartheta_{\Sel}^i)$
    \State Compute $w_{\Sel}(\vartheta_{\Sel}^i)$ by \eqref{eq:Lw}
  \EndFor
  \State $\varpi_{\Sel}^\dagger\gets\frac{\sum_{i=1}^M w_{\Sel}(\vartheta_{\Sel}^i)\vartheta_{\Sel}^i}{\sum_{i=1}^M w_{\Sel}(\vartheta_{\Sel}^i)}$
  \State Apply $u^\dagger(0)=P_{\Sel}(0)\varpi_{\Sel}^\dagger$ to the system
  \State Update $\hat\varpi_{\Sel}$
\EndFor
\end{algorithmic}
\end{algorithm}
\smallskip
\begin{remark}
  After applying the first input, Laguerre-MPPI shifts the computed input sequence by one step, projects it onto the subspace spanned by the selected basis functions by least squares, and uses the resulting coefficient as the initial estimate coefficient $\hat\varpi_{\Sel}$ at the next time.
  In conventional MPPI, the computed input sequence is shifted by one step, the last input is set to zero, and the result is used as the mean $\hat U$ of the initial estimate distribution at the next time step.
\end{remark}
\smallskip
\tabref{tab:compare} summarizes the correspondence between conventional MPPI and Laguerre-MPPI in terms of the random variables, the search dimensions, and other quantities.
\begin{table*}[t]
\centering
\caption{Correspondence between MPPI and Laguerre-MPPI}
\label{tab:compare}
\renewcommand{\arraystretch}{1.2}
\begin{tabular}{lll}
\hline
 & MPPI & Laguerre-MPPI\\
\hline
Random Variable & Input sequence $V\in\mathbb{R}^{m\times T}$ & Selected coefficient $\vartheta_{\Sel}\in\mathbb{R}^{md}$ \\
Search Dimension & $mT$ & $md$ \\
Decision Variable & $U$ & $\varpi_{\Sel}$ \\
Covariance & $\Sigma_v$ & $\Sigma_{\Sel}$ \\
Input Cost & $\frac{\lambda}{2}\sum_{k=0}^{T-1}\|u(k)-\tilde{u}(k)\|_{\Sigma_v^{-1}}^2$ & $\frac{\lambda}{2}\|\varpi_{\Sel}-\tilde\varpi_{\Sel}\|_{\Sigma_{\Sel}^{-1}}^2$ \\
Optimal Distribution & \eqref{eq:optimal_control_distribution} & \eqref{eq:qstar}\\
Optimal Mean & $\E_{\mathbb{Q}^*}[V]$ & $\E_{\mathbb{Q}_{\Sel}^*}[\vartheta_{\Sel}]$ \\
Weight & \eqref{eq:w} & \eqref{eq:Lw} \\
\hline
\end{tabular}
\end{table*}

For clarity, \tabref{tab:dist} summarizes the roles of the distributions in Laguerre-MPPI. The distributions in MPPI play the same roles.
\begin{table*}[t]
\centering
\caption{Distributions in Laguerre-MPPI}
\label{tab:dist}
\footnotesize\setlength{\tabcolsep}{3pt}\renewcommand{\arraystretch}{1.2}
\begin{tabular}{>{\raggedright\arraybackslash}p{1.55cm}>{\raggedright\arraybackslash}p{1.95cm}>{\raggedright\arraybackslash}p{6.0cm}}
\hline
Distribution & Definition & Role \\
\hline
Control $\mathbb{Q}_{\varpi_{\Sel}}$ & $\mathcal{N}(\varpi_{\Sel},\Sigma_{\Sel})$ & Its mean, the mean coefficient $\varpi_{\Sel}$, is the decision variable of Problem~\ref{prob:coef_SOC}. It is the target distribution for optimization. \\
\hline
Reference $\tilde{\mathbb{Q}}_{\Sel}$ & $\mathcal{N}(\tilde\varpi_{\Sel},\Sigma_{\Sel})$ & The distribution referenced by the control distribution $\mathbb{Q}_{\varpi_{\Sel}}$. It is often centered at zero.\\
\hline
Initial estimate $\hat{\mathbb{Q}}_{\Sel}$ & $\mathcal{N}(\hat\varpi_{\Sel},\Sigma_{\Sel})$ & Generates the samples $\vartheta_{\Sel}^i$ as the proposal distribution of importance sampling instead of sampling from the optimal distribution $\mathbb{Q}^*_{\Sel}$. \\
\hline
Optimal $\mathbb{Q}_{\Sel}^*$ & \eqref{eq:qstar} & Unique solution of the relaxed problem $\min_{\mathbb{Q}_{\Sel}}J(\mathbb{Q}_{\Sel})$. This distribution is approximated by the control distribution $\mathbb{Q}_{\varpi_{\Sel}}$ in Laguerre-MPPI. Typically, it has a complex shape. \\
\hline
\end{tabular}
\end{table*}

\subsection{Consistency with MPPI}\label{sec:coef_cov}
Laguerre-MPPI restricts the input of MPPI to an $md$-dimensional subspace.
When all basis functions are used with $N=T$, that is, when the selection matrix is $\Sel=I_{mT}$, this subspace coincides with the whole input space $\mathbb{R}^{mT}$.
Even in this case, Laguerre-MPPI does not reduce to MPPI for an arbitrary covariance matrix $\Sigma_{\Sel}$ of the selected coefficient, since $\Sigma_{\Sel}$ determines the following three quantities, which differ from those of MPPI in general.
\begin{enumerate}
\item \textit{Sampled input sequence}: The covariance matrix of the sampled input sequence $V$ generated from the selected coefficient by \eqref{eq:varthetaS} is determined by $\Sigma_{\Sel}$. Hence the sampled input sequence $V$ does not follow the initial estimate distribution $\hat{\mathbb{Q}}$ of MPPI in general.
\item \textit{Input costs}: The coefficient cost $b_{\Sel}(\varpi_{\Sel})$, defined in \eqref{eq:input_cost_coef}, differs from the input cost $b_k(u(k))$ of MPPI, defined in \eqref{eq:input_cost}.
\item \textit{Weights}: The weight $w_{\Sel}$ in \eqref{eq:Lw} assigned to each sampled input sequence $V$ differs from the weight $w$ in \eqref{eq:w} of MPPI.
\end{enumerate}
For Laguerre-MPPI to reduce to MPPI, the covariance matrix $\Sigma_{\Sel}$ must satisfy a specific condition.
We consider the following condition on the covariance matrix $\Sigma_{\Sel}$:
\begin{equation}\label{eq:SigmaS}
\Sigma_{\Sel}^{-1}=\sum_{k=0}^{T-1}P_{\Sel}(k)^\top\Sigma_v^{-1}P_{\Sel}(k).
\end{equation}
The following theorem shows the consistency of Laguerre-MPPI with MPPI.
\medskip
\begin{theorem}\label{thm:consistency}
Let $N=T$ and $\Sel=I_{mT}$, and let the covariance matrix $\Sigma_{\Sel}$ satisfy \eqref{eq:SigmaS}.
Then, for any pole $a\in[0,1)$, the following two statements hold.
\begin{enumerate}
\item Let the selected coefficient $\vartheta_{\Sel}$ be sampled from $\hat{\mathbb{Q}}_{\Sel}=\mathcal{N}(\hat\varpi_{\Sel},\Sigma_{\Sel})$. Then, the sampled input sequence $V$ generated from $\vartheta_{\Sel}$ by \eqref{eq:varthetaS} follows $\hat{\mathbb{Q}}=\mathcal{N}(\hat U,\Sigma_v)$.
\item The input $u^\dagger(0)=P_{\Sel}(0)\varpi^\dagger_{\Sel}$ generated by Algorithm~\ref{algorithm:lmppi} has the same distribution as the input $u^\dagger(0)$ generated by Algorithm~\ref{algorithm:mppi}.
\end{enumerate}
\end{theorem}
\medskip

To prove Theorem~\ref{thm:consistency}, we first present Lemmas~\ref{lem:bilinear} and \ref{lem:consistency}.
\smallskip
\begin{lemma}\label{lem:bilinear}
Under Assumption~\ref{assumption:TN}, the following two statements hold.
\begin{enumerate}
\item For vectors $\alpha,\gamma\in\mathbb{R}^{md}$, let
\begin{align}
\xi(k)&:=P_{\Sel}(k)\alpha,\label{eq:xi}\\
\zeta(k)&:=P_{\Sel}(k)\gamma.\label{eq:zeta}
\end{align}
Then, it holds that
\begin{equation}\label{eq:bilinear}
\alpha^\top\Bigl[\sum_{k=0}^{T-1}P_{\Sel}(k)^\top\Sigma_v^{-1}P_{\Sel}(k)\Bigr]\gamma=\sum_{k=0}^{T-1}\xi(k)^\top\Sigma_v^{-1}\zeta(k).
\end{equation}
\item The covariance matrix $\Sigma_{\Sel}$ satisfying \eqref{eq:SigmaS} is positive definite.
\end{enumerate}
\end{lemma}
\smallskip
\begin{proof}
\textit{1)}
By substituting \eqref{eq:xi} and \eqref{eq:zeta} into the right-hand side of \eqref{eq:bilinear}, we see that \eqref{eq:bilinear} holds.

\textit{2)}
Letting $\gamma=\alpha$ in \eqref{eq:bilinear}, we have
\begin{equation*}
\alpha^\top\Bigl[\sum_{k=0}^{T-1}P_{\Sel}(k)^\top\Sigma_v^{-1}P_{\Sel}(k)\Bigr]\alpha=\sum_{k=0}^{T-1}\xi(k)^\top\Sigma_v^{-1}\xi(k).
\end{equation*}
This quadratic form is nonnegative and equals zero only if $\xi(k)=P_{\Sel}(k)\alpha=0$ for all $k$, i.e., $\mathcal{Q}_{\Sel}\alpha=0$.
By Lemma~\ref{lem:P_sel}, $\mathcal{Q}_{\Sel}$ is of full column rank, and hence this quadratic form is positive definite.
This implies that $\Sigma_{\Sel}^{-1}$, defined in \eqref{eq:SigmaS}, is positive definite and completes the proof.
\end{proof}
\smallskip

Statement 2) of Lemma~\ref{lem:bilinear} guarantees that the Gaussian distribution with the covariance matrix $\Sigma_{\Sel}$ satisfying \eqref{eq:SigmaS} is well defined, and hence samples are drawn from it.
\smallskip
\begin{lemma}\label{lem:consistency}
Suppose that Assumptions~\ref{assumption:cost_decomposition}, \ref{assumption:quadratic_input_cost}, and \ref{assumption:TN} hold and that the covariance matrix $\Sigma_{\Sel}$ satisfies \eqref{eq:SigmaS}. Then, the following three statements hold.
\begin{enumerate}
\item For any mean coefficient $\varpi_{\Sel}$ and reference coefficient $\tilde\varpi_{\Sel}$, it holds that
\begin{equation}\label{eq:require}
b_{\Sel}(\varpi_{\Sel})=\sum_{k=0}^{T-1}b_k(u(k)).
\end{equation}
\item The sampled input sequence $V$, generated by \eqref{eq:varthetaS}, satisfies 
\begin{equation}\label{eq:Lw_input}
(\hat\varpi_{\Sel}-\tilde\varpi_{\Sel})^\top\Sigma_{\Sel}^{-1}\vartheta_{\Sel}
=\sum_{k=0}^{T-1}(\hat u(k)-\tilde u(k))^\top\Sigma_v^{-1}v(k).
\end{equation}
\item When $N=T$ and $\Sel=I_{mT}$, any mean input sequence $U$, reference input sequence $\tilde U$, and initial estimate input sequence $\hat U$ are expressed in the form of \eqref{eq:u}--\eqref{eq:uh}.
\end{enumerate}
\end{lemma}
\smallskip
\begin{proof}
\textit{1)}
Substituting \eqref{eq:u} and \eqref{eq:ut} into \eqref{eq:input_cost}, we see that the right-hand side of \eqref{eq:require} equals $\frac{\lambda}{2}\sum_{k=0}^{T-1}\|P_{\Sel}(k)(\varpi_{\Sel}-\tilde\varpi_{\Sel})\|_{\Sigma_v^{-1}}^2$.
When the covariance matrix $\Sigma_{\Sel}$ satisfies \eqref{eq:SigmaS}, applying 1) of Lemma~\ref{lem:bilinear} yields \eqref{eq:require}.

\textit{2)}
We apply 1) of Lemma~\ref{lem:bilinear} with $\alpha=\hat\varpi_{\Sel}-\tilde\varpi_{\Sel}$ and $\gamma=\vartheta_{\Sel}$.
Using the expansions \eqref{eq:ut} and \eqref{eq:uh}, we see that $\xi(k)$, defined in \eqref{eq:xi}, equals $\hat u(k)-\tilde u(k)$. Similarly, the expansion \eqref{eq:varthetaS} shows that $\zeta(k)$, defined in \eqref{eq:zeta}, equals $v(k)$.
Under the condition \eqref{eq:SigmaS}, the left-hand side of \eqref{eq:bilinear} equals the left-hand side of \eqref{eq:Lw_input}, which yields \eqref{eq:Lw_input}.

\textit{3)}
When the number of candidate basis functions is $N=T$ and the selection matrix is $\Sel=I_{mT}$, $\mathcal{Q}_{\Sel}$, defined in \eqref{eq:QSel}, becomes $\mathcal{Q}_{\Sel}=I_m\otimes\Phi^\top$, which is an $mT\times mT$ matrix and is invertible by Lemma~\ref{lem:P_sel}.
Hence its image is $\mathrm{Im}(\mathcal{Q}_{\Sel})=\mathbb{R}^{mT}$. For any mean input sequence $U\in\mathbb{R}^{m\times T}$, setting the mean coefficient to $\varpi_{\Sel}=\mathcal{Q}_{\Sel}^{-1}\mathrm{vec}(U^\top)$ gives \eqref{eq:u}. Likewise, \eqref{eq:ut} and \eqref{eq:uh} hold for $\tilde U$ and $\hat U$.
\end{proof}
\smallskip
With these preparations, we give the proof of Theorem~\ref{thm:consistency}.
\smallskip
\begin{proofof}{Theorem~\ref{thm:consistency}}
\textit{1)}
The matrix $P_{\Sel}(k)$, defined in \eqref{eq:vexp}, becomes $P_{\Sel}(k)=I_m\otimes L(k)^\top$. Hence, by the mixed-product property of the Kronecker product,
\begin{align*}
\sum_{k=0}^{T-1}P_{\Sel}(k)^\top\Sigma_v^{-1}P_{\Sel}(k)&=\Sigma_v^{-1}\otimes\sum_{k=0}^{T-1}L(k)L(k)^\top\\
&=\Sigma_v^{-1}\otimes\Phi\Phi^\top.
\end{align*}
By 2) of Lemma~\ref{lem:Phi}, the square matrix $\Phi$, defined in \eqref{eq:Phi}, is invertible. Hence \eqref{eq:SigmaS} gives the covariance matrix $\Sigma_{\Sel}=\Sigma_v\otimes(\Phi\Phi^\top)^{-1}$.
The sampled input vector $\mathcal{V}=\mathcal{Q}_{\Sel}\vartheta_{\Sel}$ follows a Gaussian distribution. Using the expansion \eqref{eq:uh}, its mean is $\mathcal{Q}_{\Sel}\hat\varpi_{\Sel}=\mathrm{vec}(\hat U^\top)$, and its covariance matrix is
\begin{equation*}
\mathcal{Q}_{\Sel}\Sigma_{\Sel}\mathcal{Q}_{\Sel}^\top=\Sigma_v\otimes\Phi^\top(\Phi\Phi^\top)^{-1}\Phi=\Sigma_v\otimes I_T.
\end{equation*}
This means that each column, the sampled input $v(k)$, independently follows $\mathcal{N}(\hat u(k),\Sigma_v)$, and thus the sampled input sequence $V$ follows the initial estimate distribution $\hat{\mathbb{Q}}$.

\textit{2)}
By 3) of Lemma~\ref{lem:consistency},
\eqref{eq:ut} and \eqref{eq:uh} hold for any $\tilde{u}(k),\hat{u}(k)$. Hence, by 2) of Lemma~\ref{lem:consistency}, the weight $w_{\Sel}(\vartheta_{\Sel}^i)$ for the selected coefficient $\vartheta_{\Sel}^i$ equals the weight $w(V^i)$ for the sampled input sequence $V^i$.
Multiplying $\varpi_{\Sel}^\dagger$ obtained in \algorithmref{algorithm:lmppi} by $\mathcal{Q}_{\Sel}$, defined in \eqref{eq:QSel}, yields
\begin{equation*}
\mathcal{Q}_{\Sel}\varpi_{\Sel}^\dagger=\frac{\sum_{i=1}^M w(V^i)\,\mathrm{vec}(V^{i\top})}{\sum_{i=1}^M w(V^i)}=\mathrm{vec}(U^{\dagger\top}).
\end{equation*}
Since the input $P_{\Sel}(0)\varpi_{\Sel}^\dagger$ applied in \algorithmref{algorithm:lmppi} is the input at time 0 extracted from the vector $\mathcal{Q}_{\Sel}\varpi_{\Sel}^\dagger$, it equals the input $u^\dagger(0)$ of MPPI.
\end{proofof}
\smallskip
\begin{remark}\label{rem:seed}
Theorem~\ref{thm:consistency} states the equality of $u^{\dagger}(0)$ in the sense of distribution.
Conventional MPPI draws the sampled input vector $\mathcal{V}$ directly from $\hat{\mathbb{Q}}$, whereas Laguerre-MPPI draws the selected coefficient $\vartheta_{\Sel}$ from $\hat{\mathbb{Q}}_{\Sel}$ and converts it into $\mathcal{V}=\mathcal{Q}_{\Sel}\vartheta_{\Sel}$.
As stated in 1) of Theorem~\ref{thm:consistency}, the converted $\mathcal{V}$ follows $\hat{\mathbb{Q}}$.
The conversion by $\mathcal{Q}_{\Sel}$ therefore moves each sample point without changing the distribution.
Hence the samples of Laguerre-MPPI are interpreted as samples drawn from $\hat{\mathbb{Q}}$ with a random seed different from that of conventional MPPI.
In particular, we let $a=0$ in the basis function $l_i(k)$, defined in \eqref{eq:li}.
This gives $\mathcal{Q}_{\Sel}=I_{mT}$ and $\mathcal{V}=\vartheta_{\Sel}$.
Therefore, the samples of the two algorithms are equal even up to the sample points.
\end{remark}

\subsection{Related Works}\label{sec:related}
The main features of Laguerre-MPPI include smooth input generation and decoupling of the search dimension from the prediction horizon $T$.
This subsection compares Laguerre-MPPI with existing variants of MPPI that share either of these features.

The first group of studies shapes the frequency content of the sampled input sequence to suppress chattering.
Vlahov et al.~\cite{10480553} replaced the white Gaussian noise of MPPI with colored noise to promote low-frequency components in the sampled input sequence.
LP-MPPI~\cite{LP-MPPI} applies a low-pass filter in the sampling process and removes the high-frequency components of the sampled noise.
SMPPI~\cite{Kim2022} lifts the input space and adds an input cost that penalizes rough input sequences.
These methods smooth the input, but they still sample the input at each time step in the prediction horizon.
Hence the number of sampled variables remains $mT$, and the curse of dimensionality for long horizons remains.

The second group of studies reduces the search dimension by sampling a small number of parameters and interpolating the input sequence from them.
Alvarez-Padilla et al.~\cite{Padilla} achieved real-time whole-body control of quadruped robots by sampling four knots for each joint and interpolating them with a cubic spline.
This spline representation reduces the search space and also smooths the input.
Tao et al.~\cite{Tao2026} compared cubic spline, B\'ezier, and piecewise linear parameterizations in MPPI for legged robots and reported that low-dimensional structured sampling is important for this application.
Schramm et al.~\cite{Schramm2025} parameterized the joint references by cubic Hermite splines, and MuJoCo MPC~\cite{Howell2022} represents action trajectories by spline knots.
In~\cite{Padilla}, the knots are evenly distributed over the prediction horizon, and hence the input is represented with the same resolution over the whole horizon.
Agentic MPC~\cite{Agentic2026} represented the input sequence using Fourier basis functions and implemented MPPI by sampling in the coefficient space.
This approach shares the central idea of Laguerre-MPPI: reducing the search dimension through a basis representation of the input sequence.
However, \cite{Agentic2026} did not provide a theoretical analysis for the resulting control method.

Laguerre-MPPI belongs to the second group and differs from the existing methods in three aspects.
First, the exponential decay of the Laguerre basis functions provides a finer representation of the control input near the beginning of the prediction horizon and a coarser representation toward its end,
as shown in Section~\ref{sec:basis}. This allocation of resolution is well suited to the receding-horizon framework, in which only the first input is applied to the plant.
Second, since the basis functions have low-pass characteristics for any pole $a>0$ and every sampled input sequence is a linear combination of them,
Laguerre-MPPI intrinsically yields the smooth input sequences that the first group aims at, without requiring an auxiliary filter or an additional cost term.
Third, Laguerre-MPPI admits theoretical analysis of closed-loop stability and sampling error, as presented in the next two sections.

\section{STABILITY ANALYSIS}\label{sec:stability}
This section analyzes the closed-loop system under Laguerre-MPPI for linear plants with quadratic costs.
Section~\ref{sec:prelim} sets up the analysis, and Section~\ref{sec:closedloop} represents the closed-loop system under Laguerre-MPPI as a stochastic system and gives a stability condition in the infinite-sample limit.
Section~\ref{sec:moment} derives closed forms of the weighted moments of the selected coefficient $\vartheta_{\Sel}$ and obtains the optimal distribution $\mathbb{Q}_{\Sel}^*$ and its mean, the optimal coefficient, explicitly.

\subsection{Linear-Quadratic Setting}\label{sec:prelim}
We restrict the analysis to the case where the nonlinear plant in \eqref{eq:sys} is given by the following linear time-invariant (LTI) system:
\begin{equation}
x(k+1)=Ax(k)+Bv(k),\label{eq:lin}
\end{equation}
where $A\in\mathbb{R}^{n\times n}$ and $B\in\mathbb{R}^{n\times m}$ are constant system matrices.
Furthermore, we replace the state cost \eqref{eq:state_cost} with the following quadratic cost:
\begin{equation}\label{eq:quad}
S(V)=x(T)^\top \Pi x(T)+\sum_{k=0}^{T-1}x(k)^\top W x(k),
\end{equation}
where $\Pi, W\in \mathbb{R}^{n\times n}$ are the weight matrices satisfying $\Pi\succeq0$ and $W\succeq0$. To address the regulation problem, we set the reference input as $\tilde u(k)\equiv 0$.
Since $\mathcal{Q}_{\Sel}$, defined in \eqref{eq:QSel}, is of full column rank as stated in Lemma~\ref{lem:P_sel}, the corresponding reference coefficient in \eqref{eq:ut} is $\tilde\varpi_{\Sel}=0$.

In the linear-quadratic setting, the state cost is expressed explicitly as follows.
Substituting the relation \eqref{eq:varthetaS} between the selected coefficient $\vartheta_{\Sel}$ and the sampled input $v(k)$ into the solution of \eqref{eq:lin} with the initial state $x(0)=x_0$, we have
\begin{equation}\label{eq:lin_sol}
x(k)=A^k x_0+\bar B_k\Sel\vartheta_{\Sel},
\end{equation}
where $\bar B_k \in\mathbb{R}^{n\times mN}$ is
\begin{equation}\label{eq:Bbar}
\bar{B}_k :=\sum_{j=0}^{k-1}A^{k-1-j}B(I_m\otimes L(j)^\top).
\end{equation}
Recalling $S_{\vartheta_{\Sel}}(\vartheta_{\Sel})$, defined in \eqref{eq:Stheta}, and substituting \eqref{eq:lin_sol} into \eqref{eq:quad}, we have the expression
\begin{equation}\label{eq:Squad}
S_{\vartheta_{\Sel}}(\vartheta_{\Sel})=\vartheta_{\Sel}^\top H_{\Sel}\vartheta_{\Sel}+2(G_{\Sel}x_0)^\top\vartheta_{\Sel},
\end{equation}
where $H_{\Sel}\in \mathbb{R}^{md\times md}$ and $G_{\Sel}\in \mathbb{R}^{md\times n}$ are given by
\begin{align}
H_{\Sel}&:=\Sel^\top H\Sel,\label{eq:HS_def}\\
G_{\Sel}&:=\Sel^\top G,\label{eq:GS_def}
\end{align}
and $H\in\mathbb{R}^{mN\times mN}$ and $G\in\mathbb{R}^{mN\times n}$ are defined as
\begin{align}
H&:=\bar B_T^\top \Pi \bar B_T+\sum_{k=0}^{T-1}\bar B_k^\top W\bar B_k\succeq0,\label{eq:H}\\
G&:=\bar B_T^\top \Pi A^{T}+\sum_{k=0}^{T-1}\bar B_k^\top W A^k.\notag
\end{align}
The constant term in \eqref{eq:quad} is omitted in \eqref{eq:Squad} since it does not affect the optimization of $\vartheta_{\Sel}$.

\subsection{Closed-Loop System and Stability}\label{sec:closedloop}
As in MPC, at each time step, Laguerre-MPPI optimizes the input sequence over the prediction horizon and applies only the first element of the sequence to the plant.
Hence the analysis of the closed-loop system requires a distinction between the actual time and the time within the prediction horizon.
We denote the actual time by $k\in\{0,1,\ldots\}$ and the prediction time by $\tau\in\{0,1,\ldots,T-1\}$.
Then, the predicted state and input $\tau$ steps ahead of the actual time $k$ are denoted by $x(\tau|k)$ and $u(\tau|k)$, respectively.
We further set the initial state of the prediction to the current state, i.e., $x(0|k)=x(k)$.

Recall that the finite-sample approximation $\varpi_{\Sel}^\dagger(k)$ of the optimal coefficient $\varpi_{\Sel}^*(k)$ is computed by \algorithmref{algorithm:lmppi}.
Then, the mean input generated by Laguerre-MPPI is expressed as $u^\dagger(k)=P_{\Sel}(0)\varpi_{\Sel}^\dagger(k)$.
The following theorem characterizes the resulting closed-loop system as a stochastic system governed by state-feedback control with additive stochastic noise.
\medskip
\begin{theorem}\label{thm:closedloop}
Consider the LTI system \eqref{eq:lin} and Laguerre-MPPI under the quadratic cost \eqref{eq:quad}.
Then, the following three statements hold.
\begin{enumerate}
\item The optimal coefficient $\varpi_{\Sel}^*(k)$ is given by
\begin{equation}\label{eq:optcoef}
\varpi_{\Sel}^*(k)=-\Bigl(H_{\Sel}+\frac{\lambda}{2}\Sigma_{\Sel}^{-1}\Bigr)^{-1}G_{\Sel}x(k).
\end{equation}
\item The mean input generated by Laguerre-MPPI at time $k$ is given by
\begin{equation}\label{eq:fb}
u^\dagger(k)=-K(\Sel)x(k)+\varepsilon_M(k),
\end{equation}
where $K(\Sel)$ and $\varepsilon_M(k)$ are given by
\begin{align}
K(\Sel)&:=P_{\Sel}(0)\Bigl(H_{\Sel}+\frac{\lambda}{2}\Sigma_{\Sel}^{-1}\Bigr)^{-1}G_{\Sel},\label{eq:K}\\
\varepsilon_M(k)&:=P_{\Sel}(0)\bigl(\varpi_{\Sel}^\dagger(k)-\varpi_{\Sel}^*(k)\bigr),\label{eq:eps}
\end{align}
respectively. 
\item For any state $x(k)$ and any initial estimate coefficient $\hat\varpi_{\Sel}(k)$, $\varepsilon_M(k) \to 0$ almost surely as $M\to\infty$.
\end{enumerate}
\end{theorem}
\medskip
The proof of Theorem~\ref{thm:closedloop} is given in Section~\ref{sec:moment}.
\begin{figure*}[t]
\centering
\figorbox{0.8\linewidth}{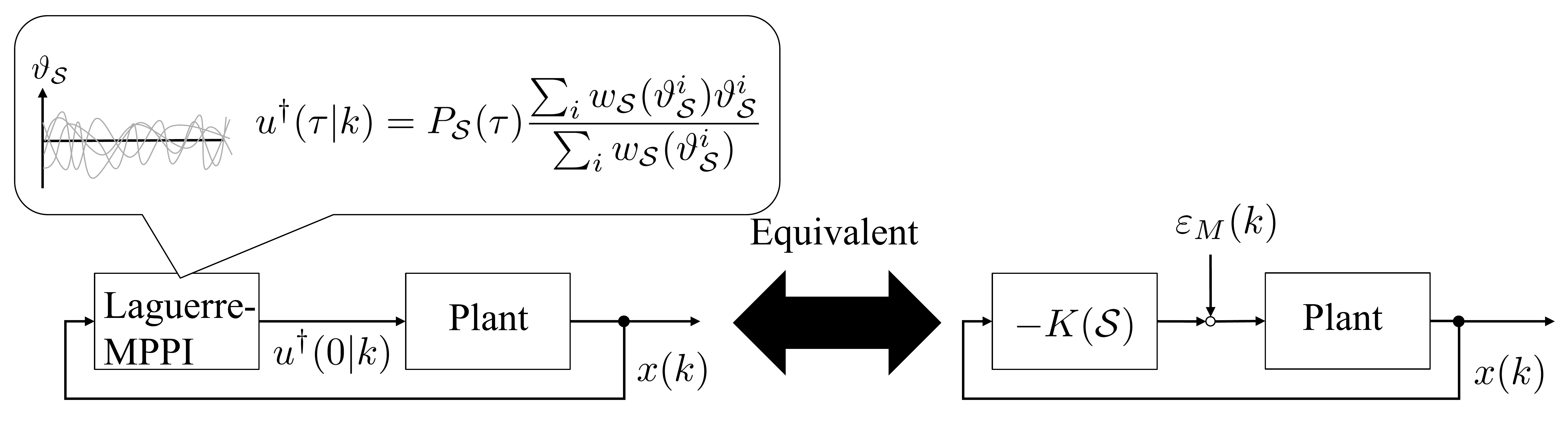}
\caption{Closed-loop system under Laguerre-MPPI (left) and its equivalent representation given by Theorem~\ref{thm:closedloop} (right).}
\label{fig:closedloop}
\end{figure*}

The finite-sample approximation $\varpi_{\Sel}^\dagger(k)$ in Theorem~\ref{thm:closedloop} determines the predicted input sequence within the prediction horizon,
\begin{equation}\label{eq:u_pred}
u^\dagger(\tau|k)=P_{\Sel}(\tau)\varpi_{\Sel}^\dagger(k),\quad \tau\in\{0,1,\ldots,T-1\},
\end{equation}
where $P_{\Sel}(\tau)$ is defined in \eqref{eq:vexp}.
Laguerre-MPPI applies only the first input $u^\dagger(k)=u^\dagger(0|k)$ of \eqref{eq:u_pred}. As in MPC, the state is updated to $x(k+1)$,
the prediction horizon is shifted by one step, and the finite-sample approximation $\varpi_{\Sel}^\dagger(k+1)$ is computed with the state $x(k+1)$ as the initial state.

Statement 2) shows that the first input $u^\dagger(0|k)$ of \eqref{eq:u_pred} is equivalently rewritten in the feedback form \eqref{eq:fb}, as illustrated in \figref{fig:closedloop}.
The feedback gain $K(\Sel)$ in the first term of \eqref{eq:fb} is determined only by the covariance matrix $\Sigma_{\Sel}$, the temperature parameter $\lambda$,
and the matrices $P_{\Sel}(0)$, $H_{\Sel}$, and $G_{\Sel}$, defined in \eqref{eq:vexp}, \eqref{eq:HS_def}, and \eqref{eq:GS_def}, respectively.
Hence, $K(\Sel)$ is time-invariant.
The second term, the stochastic noise $\varepsilon_M(k)$, defined in \eqref{eq:eps}, arises from the difference between the finite-sample approximation $\varpi_{\Sel}^\dagger(k)$ and the optimal coefficient $\varpi_{\Sel}^*(k)$.
It varies randomly with the generated samples.
Hence Laguerre-MPPI applies the linear time-invariant state feedback $-K(\Sel)x(k)$ with additive stochastic noise. Substituting \eqref{eq:fb} into \eqref{eq:lin},
we see that the closed-loop system is the stochastic system
\begin{equation}\label{eq:stoch}
x(k+1)=\bigl(A-BK(\Sel)\bigr)x(k)+B\varepsilon_M(k).
\end{equation}

Statement 3) shows that the stochastic noise $\varepsilon_M(k)$ converges to zero as $M\to\infty$, and the closed-loop system becomes
\begin{equation}\label{eq:nominal}
x(k+1)=\bigl(A-BK(\Sel)\bigr)x(k).
\end{equation}
Thus, infinite-sample Laguerre-MPPI reduces to a linear time-invariant state feedback law.
The following corollary presents the condition for closed-loop stability in this limit.
\smallskip
\begin{corollary}\label{col:stab}
Consider the control system composed of \eqref{eq:lin} and Laguerre-MPPI with the quadratic cost \eqref{eq:quad} in the infinite-sample limit.
Then, the origin is asymptotically stable if and only if $A-BK(\Sel)$ is Schur stable, i.e.,
\begin{equation}\label{eq:rho}
\rho\bigl(A-BK(\Sel)\bigr)<1.
\end{equation}
\end{corollary}
\smallskip
\begin{remark}\label{rem:mpc}
For the LTI system \eqref{eq:lin} with the quadratic cost \eqref{eq:quad}, the expected state cost satisfies $\E_{\mathbb{Q}_{\varpi_{\Sel}}}[S_{\vartheta_{\Sel}}(\vartheta_{\Sel})]=S_{\vartheta_{\Sel}}(\varpi_{\Sel})+\mathrm{tr}(H_{\Sel}\Sigma_{\Sel})$. Hence the minimizer of \eqref{eq:SOC_coef} is given by \eqref{eq:optcoef}.
In particular, when $\Sigma_{\Sel}$ satisfies \eqref{eq:SigmaS}, 1) of Lemma~\ref{lem:consistency} shows that the coefficient cost $b_{\Sel}(\varpi_{\Sel})$ becomes $\frac{\lambda}{2}\sum_k u(k)^\top\Sigma_v^{-1}u(k)$.
In this case, infinite-sample Laguerre-MPPI reduces to an MPC that solves the finite-horizon LQ problem with input weight $\frac{\lambda}{2}\Sigma_v^{-1}$ restricted to the subspace spanned by the selected basis functions, and it has the same structure as Laguerre-MPC~\cite{Wang2001}.
\end{remark}
\smallskip
Note that the control law under Laguerre-MPPI, given by \eqref{eq:fb}, depends on the selection matrix $\Sel$. Consequently, the closed-loop stability can be lost depending on the basis selection.
Section~\ref{sec:exp3} confirms this dependence with a numerical example.

\subsection{Weighted Moments and Proof of Theorem~\ref{thm:closedloop}}\label{sec:moment}
This subsection presents the proof of Theorem~\ref{thm:closedloop}, stated in Section~\ref{sec:closedloop}. As preparation, we derive the weighted moments of the selected coefficient $\vartheta_{\Sel}$ with respect to the initial estimate distribution $\hat{\mathbb{Q}}_{\Sel}$.
Since $S_{\vartheta_{\Sel}}(\vartheta_{\Sel})$ in \eqref{eq:Squad} is a quadratic form in $\vartheta_{\Sel}$, the product $w_{\Sel}^p\hat{q}_{\Sel}$ of the $p$-th power of the weight \eqref{eq:Lw} and the density $\hat{q}_{\Sel}$ of $\hat{\mathbb{Q}}_{\Sel}$ is proportional to a Gaussian probability density for any $p>0$.
To describe this Gaussian density, we define, for $p>0$,
\begin{gather}
\Lambda_p:=\Sigma_{\Sel}^{-1}+\frac{2p}{\lambda}H_{\Sel},\label{eq:Lambda_p}\\
b_p:=\frac{2p}{\lambda}G_{\Sel}x_0-\Sigma_{\Sel}^{-1}\bigl[(1-p)\hat\varpi_{\Sel}+p\tilde\varpi_{\Sel}\bigr],\label{eq:b_p}\\
m_p:=-\Lambda_p^{-1}b_p,\label{eq:m_p}\\
\mathbb{Q}_p:=\mathcal{N}(m_p,\Lambda_p^{-1}).\label{eq:Qp}
\end{gather}
In addition, for simplicity of notation, we let
\begin{equation}\label{eq:Zdef}
  Z_p:=\E_{\hat{\mathbb{Q}}_{\Sel}}[w_{\Sel}^p].
\end{equation}
The following lemma shows that dividing the product $w_{\Sel}^p\hat{q}_{\Sel}$ by $Z_p$ yields the density $q_p$ of the Gaussian distribution $\mathbb{Q}_p$.
Hence $m_p$ and $\Lambda_p^{-1}$ are the mean and the covariance matrix of the normalized product $w_{\Sel}^p\hat{q}_{\Sel}/Z_p$, respectively.
In particular, for $p=1$, the Gaussian distribution $\mathbb{Q}_1$ equals the optimal distribution $\mathbb{Q}_{\Sel}^*$.
The vector $-b_p$ is the coefficient vector of the linear term in the exponent of $w_{\Sel}^p\hat{q}_{\Sel}$.
We use the case $p=1$ for the optimal coefficient in Theorem~\ref{thm:closedloop} and the case $p=2$ for the ESS in Section~\ref{sec:ess}.
\smallskip
\begin{lemma}\label{lem:moment}
For $p>0$, let $\Lambda_p$, $b_p$, $m_p$, $\mathbb{Q}_p$, and $Z_p$ be defined by \eqref{eq:Lambda_p}--\eqref{eq:Zdef}.
Then, the following five statements hold.
\begin{enumerate}
\item $Z_p$ is given by
\begin{equation}\label{eq:Zp}
Z_p=\frac{\exp\bigl(\frac12b_p^\top\Lambda_p^{-1}b_p-\frac12\hat\varpi_{\Sel}^\top\Sigma_{\Sel}^{-1}\hat\varpi_{\Sel}\bigr)}{|\Sigma_{\Sel}|^{1/2}|\Lambda_p|^{1/2}},
\end{equation}
and $Z_p \in (0,\infty)$ holds.
\item It holds that
\begin{equation}\label{eq:wq_density}
q_p(\vartheta_{\Sel})=\frac{w_{\Sel}^p(\vartheta_{\Sel})\,\hat{q}_{\Sel}(\vartheta_{\Sel})}{Z_p},
\end{equation}
for any selected coefficient $\vartheta_{\Sel}\in\mathbb{R}^{md}$.
In particular, $\mathbb{Q}_1=\mathbb{Q}_{\Sel}^*$, whose density is given by \eqref{eq:qstar}.
\item For any measurable function $\varphi$ satisfying $\E_{\mathbb{Q}_p}[|\varphi(\vartheta_{\Sel})|]<\infty$,
\begin{equation}\label{eq:change_measure}
\E_{\mathbb{Q}_p}[\varphi(\vartheta_{\Sel})]=\frac{\E_{\hat{\mathbb{Q}}_{\Sel}}[w_{\Sel}^p\varphi(\vartheta_{\Sel})]}{Z_p}.
\end{equation}
\item It holds that
\begin{gather}
m_p=\frac{\E_{\hat{\mathbb{Q}}_{\Sel}}[w_{\Sel}^p\vartheta_{\Sel}]}{Z_p},\label{eq:moment1}\\
\Lambda_p^{-1}+m_pm_p^\top=\frac{\E_{\hat{\mathbb{Q}}_{\Sel}}[w_{\Sel}^p\vartheta_{\Sel}\vartheta_{\Sel}^\top]}{Z_p}.\label{eq:moment2}
\end{gather}
\item For any $r\ge0$,
\begin{equation}\label{eq:moment_r}
\E_{\hat{\mathbb{Q}}_{\Sel}}[w_{\Sel}^p\|\vartheta_{\Sel}\|^r]<\infty.
\end{equation}
\end{enumerate}
\end{lemma}
\smallskip
\begin{proof}
See Appendix~\ref{app:proof_lem_moment}.
\end{proof}
\smallskip
Based on Lemma~\ref{lem:moment}, we give the proof of Theorem~\ref{thm:closedloop}.
Recall that Lemma~\ref{lem:moment} holds for the initial state $x_0$ and the initial estimate coefficient $\hat\varpi_{\Sel}$.
In the following proof of the theorem, we use the lemma with $x_0$ and $\hat\varpi_{\Sel}$ replaced by $x(k)$ and $\hat\varpi_{\Sel}(k)$, respectively, since the predicted state satisfies $x(0|k)=x(k)$.

\smallskip
\begin{proofof}{Theorem~\ref{thm:closedloop}}
\textit{1)}
Applying \eqref{eq:moment1} with $p=1$ and the definition \eqref{eq:Zdef} of $Z_1$ to the numerator and denominator of \eqref{eq:LIS} gives $\varpi_{\Sel}^*(k)=Z_1m_1/Z_1=m_1$.
Since the reference coefficient is $\tilde\varpi_{\Sel}=0$, \eqref{eq:Lambda_p} and \eqref{eq:m_p} give
\begin{equation*}
\varpi_{\Sel}^*(k)=m_1=-\Bigl(H_{\Sel}+\frac{\lambda}{2}\Sigma_{\Sel}^{-1}\Bigr)^{-1}G_{\Sel}x(k),
\end{equation*}
which equals \eqref{eq:optcoef}.

\textit{2)}
By \algorithmref{algorithm:lmppi}, the input applied at time $k$ is $u^\dagger(k)=P_{\Sel}(0)\varpi_{\Sel}^\dagger(k)$. Substituting $\varpi_{\Sel}^\dagger(k)=\varpi_{\Sel}^*(k)+\bigl(\varpi_{\Sel}^\dagger(k)-\varpi_{\Sel}^*(k)\bigr)$ gives
\begin{equation*}
u^\dagger(k)=P_{\Sel}(0)\varpi_{\Sel}^*(k)+P_{\Sel}(0)\bigl(\varpi_{\Sel}^\dagger(k)-\varpi_{\Sel}^*(k)\bigr).
\end{equation*}
The first term equals $-K(\Sel)x(k)$, and the second term is the stochastic noise $\varepsilon_M(k)$, defined in \eqref{eq:eps}. Hence \eqref{eq:fb} follows.

\textit{3)}
This statement is the consistency of self-normalized importance sampling~\cite{geweke1989bayesian,Owen2013}. We show that the stochastic noise $\varepsilon_M(k)$ converges to zero almost surely as $M\to\infty$.
The samples $\vartheta_{\Sel}^1,\ldots,\vartheta_{\Sel}^M$ are generated independently from the initial estimate distribution $\hat{\mathbb{Q}}_{\Sel}$.
We write the finite-sample approximation $\varpi_{\Sel}^\dagger(k)$ in \algorithmref{algorithm:lmppi} as the ratio of the sample means of the numerator and the denominator,
\begin{equation}\label{eq:ND}
\varpi_{\Sel}^\dagger(k)=\frac{\bar N_M}{\bar D_M},
\end{equation}
where $\bar{N}_M$ and $\bar D_M$ are given by
\begin{align*}
\bar N_M:=&\frac1M\sum_{i=1}^Mw_{\Sel}(\vartheta_{\Sel}^i)\,\vartheta_{\Sel}^i,\\
\bar D_M:=&\frac1M\sum_{i=1}^Mw_{\Sel}(\vartheta_{\Sel}^i),
\end{align*}
respectively.
Since $w_{\Sel}$ and $\bar D_M$ are positive, \eqref{eq:ND} is always well defined.

By 5) of Lemma~\ref{lem:moment}, $\E_{\hat{\mathbb{Q}}_{\Sel}}[w_{\Sel}]<\infty$ and $\E_{\hat{\mathbb{Q}}_{\Sel}}[w_{\Sel}\|\vartheta_{\Sel}\|]<\infty$.
Hence the weights $w_{\Sel}(\vartheta_{\Sel}^i)$ and each entry of $w_{\Sel}(\vartheta_{\Sel}^i)\vartheta_{\Sel}^i$ are integrable independent and identically distributed random variables.
As $M\to\infty$, the strong law of large numbers, the definition \eqref{eq:Zdef} of $Z_1$, and \eqref{eq:moment1} yield
\begin{align}
\bar N_M&\to\E_{\hat{\mathbb{Q}}_{\Sel}}[w_{\Sel}\vartheta_{\Sel}]=Z_1m_1\quad\text{a.s.,}\notag\\
\bar D_M&\to\E_{\hat{\mathbb{Q}}_{\Sel}}[w_{\Sel}]=Z_1\quad\text{a.s.}\label{eq:ND_lim}
\end{align}
Since $Z_1$, defined in \eqref{eq:Zdef}, is positive by 1) of Lemma~\ref{lem:moment}, the ratio map $(\bar N_M,\bar D_M)\mapsto\bar N_M/\bar D_M$ is continuous at the point $(Z_1m_1,Z_1)$.
Hence, combining \eqref{eq:ND} with the limits \eqref{eq:ND_lim}, we see that the finite-sample approximation $\varpi_{\Sel}^\dagger(k)$ converges almost surely to the optimal coefficient $m_1=\varpi_{\Sel}^*(k)$.
Since the stochastic noise $\varepsilon_M(k)$, defined in \eqref{eq:eps}, is a linear map of the sampling error $\varpi_{\Sel}^\dagger(k)-\varpi_{\Sel}^*(k)$, it satisfies $\varepsilon_M(k)\to0$ a.s.
\end{proofof}
\smallskip

\section{SAMPLING EFFICIENCY ANALYSIS}\label{sec:ess}
For a finite number of samples $M$, the stochastic noise $\varepsilon_M$, defined in \eqref{eq:eps}, perturbs the closed-loop dynamics as described by \eqref{eq:stoch}.
This section derives estimates of the asymptotic covariance of the sampling error $\varpi_{\Sel}^\dagger-\varpi_{\Sel}^*$ and the asymptotic ESS ratio, and shows that they are characterized by a single quantity.
\subsection{Effective Sample Size}\label{sec:essdef}
To assess sampling efficiency, we examine the concentration of the importance weights.
As the weights become concentrated on fewer samples, the number of samples that effectively contribute to the finite-sample approximation decreases.
We measure the sampling efficiency by the ESS~\cite{Kong1992,agapiou2017importance}, defined by
\begin{equation}\label{eq:ESS}
\ESS:=\frac{\bigl(\sum_{i=1}^Mw_{\Sel}(\vartheta_{\Sel}^i)\bigr)^2}{\sum_{i=1}^Mw_{\Sel}(\vartheta_{\Sel}^i)^2}.
\end{equation}
The ESS takes values in the range $1\le\ESS\le M$.
It equals $M$ when all weights are equal, whereas it equals $1$ when the weights concentrate on a single sample. Hence, a larger ESS implies a higher sampling efficiency.
The ESS approximately represents the number of independent samples from the optimal distribution $\mathbb{Q}_{\Sel}^*$ to which the $M$ weighted samples are equivalent~\cite{Kong1992,geweke1989bayesian}.
In the following, we fix the state $x(k)$ and the initial estimate coefficient $\hat\varpi_{\Sel}(k)$ at time $k$, and denote the difference between $\hat{\varpi}_{\Sel}$ and the optimal coefficient $\varpi_{\Sel}^*$ by
\begin{equation}\label{eq:e}
\nu:=\hat\varpi_{\Sel}-\varpi_{\Sel}^*.
\end{equation}
Recalling $Z_p$ of \eqref{eq:Zdef}, we define
\begin{equation}\label{eq:kappaL}
\kappa^{\Sel}:=\frac{Z_2}{Z_1^2}.
\end{equation}
\smallskip
\begin{remark}
The quantity $\kappa^{\Sel}$ is the second moment of the normalized weight $w_{\Sel}/Z_1$ under the initial estimate distribution $\hat{\mathbb{Q}}_{\Sel}$.
A small $\kappa^{\Sel}$ is desirable since $\kappa^{\Sel}$ measures the nonuniformity of the weights.
\end{remark}
\smallskip
The following theorem is obtained by applying the asymptotic theory of self-normalized importance sampling~\cite{geweke1989bayesian,agapiou2017importance} to Laguerre-MPPI.
It shows that the asymptotic covariance of the sampling error $\varpi_{\Sel}^\dagger-\varpi_{\Sel}^*$ scales with $\kappa^{\Sel}$ and that the asymptotic ESS ratio equals $1/\kappa^{\Sel}$, and it gives $\kappa^{\Sel}$ in closed form.
\medskip
\begin{theorem}\label{thm:clt}
Consider the LTI system \eqref{eq:lin} and Laguerre-MPPI under the quadratic cost \eqref{eq:quad}, and fix the state $x(k)$ and the initial estimate coefficient $\hat\varpi_{\Sel}(k)$ at time $k$.
Then, the following three statements hold.
\begin{enumerate}
\item (Asymptotic normality)
\begin{equation*}
\sqrt{M}(\varpi_{\Sel}^\dagger-\varpi_{\Sel}^*)\xrightarrow{d}\mathcal{N}(0,\Xi(\kappa^{\Sel})), \ M \to\infty,
\end{equation*}
where $\Xi(\kappa^{\Sel})$ is given by
\begin{equation}\label{eq:Xi}
\Xi(\kappa^{\Sel}):=\kappa^{\Sel}[\Lambda_2^{-1}+(m_2-m_1)(m_2-m_1)^\top],
\end{equation}
and $\Lambda_2, m_1$ and $m_2$ are defined by \eqref{eq:Lambda_p} and \eqref{eq:m_p} with $p=1,2$.
\item (Asymptotic ESS ratio)
\begin{equation*}
\frac{\ESS}{M}\to\frac{1}{\kappa^{\Sel}}\quad \text{a.s.}, \ M \to\infty.
\end{equation*}
\item (Closed form of $\kappa^{\Sel}$)
\begin{equation}
\kappa^{\Sel}=\kappa_0^{\Sel}\exp\Bigl(\frac12\nu^\top\Psi \nu\Bigr),\label{eq:kappa}
\end{equation}
where $\Psi\succ 0$ and $\kappa_0^{\Sel}$ are given by
\begin{gather}
\Psi:=\Sigma_{\Sel}^{-1}+\Sigma_{\Sel}^{-1}\Lambda_2^{-1}\Sigma_{\Sel}^{-1},\label{eq:Psi}\\
\kappa_0^{\Sel}:=\frac{\bigl|I_{md}+\frac{2}{\lambda}\Sigma_{\Sel}H_{\Sel}\bigr|}{\bigl|I_{md}+\frac{4}{\lambda}\Sigma_{\Sel}H_{\Sel}\bigr|^{1/2}},\label{eq:kappa0}
\end{gather}
respectively, and $H_{\Sel}$ is defined in \eqref{eq:HS_def}.
\end{enumerate}
\end{theorem}
\medskip
\begin{proof}
See Appendix~\ref{app:proof_thm_clt}.
\end{proof}
\smallskip
\begin{remark}
  As shown in \eqref{eq:ND}, the finite-sample approximation $\varpi_{\Sel}^\dagger$ is the ratio of $\bar N_M$ and $\bar D_M$, which are computed from the sampled coefficients $\vartheta_{\Sel}^i$ and their weights $w_{\Sel}(\vartheta_{\Sel}^i)$.
Since the expectation of a ratio does not equal the ratio of the expectations in general, the mean and the covariance matrix of the sampling error $\varpi_{\Sel}^\dagger-\varpi_{\Sel}^*$ are difficult to obtain directly.
Rewriting $\varpi_{\Sel}^\dagger$ around the limits $Z_1m_1$ and $Z_1$ of $\bar N_M$ and $\bar D_M$ yields
\begin{equation}\label{eq:varpi_diff}
\begin{split}
\varpi_{\Sel}^\dagger-\varpi_{\Sel}^*
=&\frac{1}{M}\frac{1}{Z_1}\sum_{i=1}^M w_{\Sel}(\vartheta_{\Sel}^i)(\vartheta_{\Sel}^i-m_1)\\
&+(\bar N_M-m_1\bar D_M)\Bigl(\frac{1}{\bar D_M}-\frac{1}{Z_1}\Bigr).
\end{split}
\end{equation}
The first term is linear in $\bar N_M$ and $\bar D_M$.
This term is the sample mean of the independent random vectors $w_{\Sel}(\vartheta_{\Sel}^i)(\vartheta_{\Sel}^i-m_1)/Z_1$, and hence its mean is zero and its covariance matrix is $\Xi(\kappa^{\Sel})/M$.
In the second term, both $\bar N_M-m_1\bar D_M$ and $1/\bar D_M-1/Z_1$ are of order $1/\sqrt{M}$.
Hence the second term is of order $1/M$ and converges to zero faster than the first term of order $1/\sqrt{M}$.
From this viewpoint, statement 1) is interpreted as follows: for sufficiently large $M$, the probability that the second term is not negligible compared with the first term is small, and the distribution of $\varpi_{\Sel}^\dagger-\varpi_{\Sel}^*$ approaches $\mathcal{N}(0,\Xi(\kappa^{\Sel})/M)$ of the first term.
However, nothing is guaranteed about its mean and covariance matrix.
\end{remark}
\smallskip
  \begin{remark}
As stated in 1) and 2), a large $\kappa^{\Sel}$ reduces the asymptotic ESS $M/\kappa^{\Sel}$ and enlarges the asymptotic covariance matrix $P_{\Sel}(0)\Xi(\kappa^{\Sel})P_{\Sel}(0)^\top/M$ of $\varepsilon_M(k)$, defined in \eqref{eq:eps}.
Hence a larger $\kappa^{\Sel}$ requires a larger number of samples $M$ for the closed-loop system \eqref{eq:stoch} to approximate the infinite-sample limit \eqref{eq:nominal}.
Statement 3) shows what makes $\kappa^{\Sel}$ large.
As shown in \eqref{eq:kappa}, $\kappa^{\Sel}$ is the product of the two factors $\kappa_0^{\Sel}$ and $\exp(\frac12\nu^\top\Psi\nu)$. These two factors are interpreted as follows.
\begin{enumerate}
\item The factor $\kappa_0^{\Sel}$, defined in \eqref{eq:kappa0}, represents the reduction of the ESS caused by the difference in covariance between the initial estimate distribution $\hat{\mathbb{Q}}_{\Sel}$ and the optimal distribution $\mathbb{Q}_{\Sel}^*$.
It equals $\kappa^{\Sel}$ for $\nu=0$.
Even in this case, the two distributions differ in covariance unless $H_{\Sel}=0$.
As stated in 2) of Lemma~\ref{lem:moment}, the covariance matrix of $\mathbb{Q}_{\Sel}^*=\mathbb{Q}_1$ is $\Lambda_1^{-1}$, and $\Lambda_1\succeq\Sigma_{\Sel}^{-1}$ by \eqref{eq:Lambda_p} implies $\Lambda_1^{-1}\preceq\Sigma_{\Sel}$.
Hence the samples drawn from $\hat{\mathbb{Q}}_{\Sel}$ spread more widely than those from $\mathbb{Q}_{\Sel}^*$, and the samples far from $\varpi_{\Sel}^*$ receive small weights.
Appendix~\ref{app:details_kappa0} shows the details.
\item The factor $\exp(\frac12\nu^\top\Psi\nu)$ represents the reduction of the ESS caused by the difference $\nu$ between the means.
When $\hat\varpi_{\Sel}$ deviates from $\varpi_{\Sel}^*$, most samples lie away from the region where the optimal distribution has a large density, and the weights concentrate on the few samples near $\varpi_{\Sel}^*$.
Since $\Psi$, defined in \eqref{eq:Psi}, is positive definite, this factor is no less than 1, equals 1 only for $\nu=0$, and grows exponentially in $\nu^\top\Psi\nu$.
\end{enumerate}
\end{remark}
\smallskip
To relate the number of samples to the error from the infinite-sample limit, we impose the following assumption on the stochastic noise $\varepsilon_M(k)$.
\smallskip
\begin{assumption}\label{ass:eps}
The stochastic noise $\varepsilon_M(k)$, defined in \eqref{eq:eps}, satisfies
\begin{equation}\label{eq:eps_mom}
\begin{aligned}
\E[\varepsilon_M(k)\mid x(k),\hat\varpi_{\Sel}]&= 0,\\
\mathrm{Cov}[\varepsilon_M(k) \mid x(k),\hat\varpi_{\Sel}]&=\frac{1}{M}P_{\Sel}(0)\Xi(\kappa^{\Sel})P_{\Sel}(0)^\top.
\end{aligned}
\end{equation}
\end{assumption}
\smallskip
Assumption \eqref{eq:eps_mom} replaces the mean and the covariance matrix of the sampling error $\varpi_{\Sel}^\dagger-\varpi_{\Sel}^*$ by those of the first term of \eqref{eq:varpi_diff}.
This is interpreted as assuming that the rare events in which the second term of \eqref{eq:varpi_diff} becomes large contribute negligibly to the mean and the covariance matrix.

Moreover, we let $x_{\mathrm{nom}}(k)$ and $\chi(k):=x(k)-x_{\mathrm{nom}}(k)$ denote the state of the infinite-sample limit \eqref{eq:nominal} with $x_{\mathrm{nom}}(0)=x(0)$ and the error between the actual state $x(k)$ of the closed-loop system \eqref{eq:stoch} and $x_{\mathrm{nom}}(k)$, respectively.
The following corollary gives a necessary condition on $M$ for the root-mean-square error (RMSE) $\E[\|\chi(k)\|^2]^{1/2}$ to stay within a tolerance $\chi_{\mathrm{tol}}>0$.
\smallskip
\begin{corollary}\label{cor:error}
Suppose that $A-BK(\Sel)$ is Schur stable and that Assumption~\ref{ass:eps} holds.
Let
\begin{equation}\label{eq:Sigma_eps}
\Sigma_\varepsilon:=\kappa_0^{\Sel}P_{\Sel}(0)\Lambda_2^{-1}P_{\Sel}(0)^\top,
\end{equation}
and $\Upsilon$ be the unique solution of the Lyapunov equation
\begin{equation}\label{eq:Lyap_eq}
\Upsilon=\bigl(A-BK(\Sel)\bigr)\Upsilon\bigl(A-BK(\Sel)\bigr)^\top+B\Sigma_\varepsilon B^\top.
\end{equation}
If $\E[\|\chi(k)\|^2]\le\chi_{\mathrm{tol}}^2$ holds for all $k\ge0$, then
\begin{equation}\label{eq:Mreq}
M\ge\frac{\mathrm{tr}\,\Upsilon}{\chi_{\mathrm{tol}}^2}.
\end{equation}
\end{corollary}
\begin{proof}
  See Appendix~\ref{app:error}.
\end{proof}
\smallskip
This shows that if $M$ does not satisfy \eqref{eq:Mreq}, the RMSE $\E[\|\chi(k)\|^2]^{1/2}$ exceeds $\chi_{\mathrm{tol}}$ at some time $k$.
Hence, Corollary~\ref{cor:error} provides a guideline on the number of samples required to accurately approximate the stochastic system \eqref{eq:stoch} by the infinite-sample limit \eqref{eq:nominal}.
Note that even if $M$ satisfies \eqref{eq:Mreq}, the RMSE may exceed $\chi_{\mathrm{tol}}$ at some time $k$.
Since $\Sigma_\varepsilon$ and $\Upsilon$ depend only on the design parameters, this lower bound on $M$ is evaluated before the samples are generated.

\subsection{Search Dimension Reduction and the ESS Bound}\label{sec:dim}
As stated in 2) of Theorem~\ref{thm:clt}, the asymptotic ESS ratio equals $1/\kappa^{\Sel}$.
Hence the quantity $\kappa^{\Sel}$, defined in \eqref{eq:kappaL}, indicates the sampling efficiency, and a smaller $\kappa^{\Sel}$ is desirable.
As stated in 3) of Theorem~\ref{thm:clt}, $\kappa^{\Sel}$ attains its minimum $\kappa_0^{\Sel}$ when the initial estimate coefficient $\hat\varpi_{\Sel}$ equals the optimal coefficient $\varpi_{\Sel}^*$.
This subsection shows that $\kappa_0^{\Sel}$ of Laguerre-MPPI is less than that of conventional MPPI, which implies the improved sampling efficiency of Laguerre-MPPI.
In this subsection, the covariance matrix $\Sigma_{\Sel}$ satisfies \eqref{eq:SigmaS}.
Then, for fixed $T$, $a$, $\lambda$, $\Sigma_v$, and the system and cost matrices, $\kappa_0^{\Sel}$ is determined by the number $N$ of candidate basis functions and the selection matrix $\Sel$.
To make this dependence explicit, we write
\begin{equation*}
\kappa_0(N,\Sel):=\kappa_0^{\Sel}.
\end{equation*}
Then, the following theorem holds.
\medskip
\begin{theorem}\label{thm:kappa0}
Suppose that Assumption~\ref{assumption:TN} holds and that the covariance matrix $\Sigma_{\Sel}$ is given by \eqref{eq:SigmaS}.
Let $\mu_1(N,\Sel)\ge\cdots\ge\mu_{md}(N,\Sel)$ be the eigenvalues of the symmetric positive semidefinite matrix $\Sigma_{\Sel}^{1/2}H_{\Sel}\Sigma_{\Sel}^{1/2}$,
where $\Sigma_{\Sel}^{1/2}$ is the symmetric square root of $\Sigma_{\Sel}$, and let the function $\psi$ be
\begin{equation}\label{eq:psi}
\psi(\mu):=\frac{1+2\mu/\lambda}{\sqrt{1+4\mu/\lambda}}.
\end{equation}
Then, the following two statements hold.
\begin{enumerate}
\item $\kappa_0(N,\Sel)$ is given by
\begin{equation}\label{eq:kappa0_prod}
\kappa_0(N,\Sel)=\prod_{i=1}^{md}\psi\bigl(\mu_i(N,\Sel)\bigr).
\end{equation}
\item For any selection matrix $\Sel$,
\begin{equation}\label{eq:kappa0_ineq}
\kappa_0(N,\Sel)\le\kappa_0(T,I_{mT}).
\end{equation}
\end{enumerate}
\end{theorem}
\medskip
\begin{proof}
See Appendix~\ref{app:proof_thm_kappa0}.
\end{proof}
\smallskip
\begin{remark}\label{rem:kappa0_mppi}
Recall from Theorem~\ref{thm:consistency} that Laguerre-MPPI with $N=T$, $\Sel=I_{mT}$, and the covariance matrix satisfying \eqref{eq:SigmaS} generates the sampled input sequence $V$ following the initial estimate distribution $\hat{\mathbb{Q}}$ of conventional MPPI for any pole $a$.
In addition, as stated in 2) and 3) of Lemma~\ref{lem:consistency}, its weight $w_{\Sel}$ equals the weight $w$ of conventional MPPI, defined in \eqref{eq:w}, evaluated at $V$.
Hence, Laguerre-MPPI with these conditions reduce to conventional MPPI, and $\kappa_0(T,I_{mT})$ is the quantity corresponding to conventional MPPI.
\end{remark}
\smallskip
As stated in 2) and 3) of Theorem~\ref{thm:clt}, the asymptotic ESS ratio of Laguerre-MPPI is
\begin{equation*}
\lim_{M\to\infty}\frac{\ESS}{M}=\frac{1}{\kappa_0(N,\Sel)}\exp\Bigl(-\frac12\nu^\top\Psi\nu\Bigr)\le\frac{1}{\kappa_0(N,\Sel)}\quad\text{a.s.},
\end{equation*}
where the equality holds if and only if $\nu=0$, i.e., the initial estimate coefficient $\hat\varpi_{\Sel}$ equals the optimal coefficient $\varpi_{\Sel}^*$.
Hence $1/\kappa_0(N,\Sel)$ is the maximum of the asymptotic ESS ratio over $\hat\varpi_{\Sel}$.
Moreover, 2) of Theorem~\ref{thm:kappa0} gives
\begin{equation*}
\frac{1}{\kappa_0(T,I_{mT})}\le\frac{1}{\kappa_0(N,\Sel)}.
\end{equation*}
The left-hand side is the maximum for conventional MPPI.
Therefore, for any pole $a$ and any basis selection, the maximum of the asymptotic ESS ratio of Laguerre-MPPI is no less than that of conventional MPPI.
This result quantifies the effect of the search dimension reduction, which keeps the samples dense in the coefficient space.
Note that \eqref{eq:kappa0_ineq} compares the maxima attained at $\nu=0$ and does not imply the same ordering for each $\hat\varpi_{\Sel}$.

\section{NUMERICAL EXPERIMENTS}\label{sec:experiment}
Sections~\ref{sec:exp1} and \ref{sec:exp2} demonstrate through numerical experiments that Laguerre-MPPI produces smooth inputs and that its control performance is robust to the extension of the prediction horizon.
Section~\ref{sec:exp2} also verifies numerically the inequality \eqref{eq:kappa0_ineq} in Theorem~\ref{thm:kappa0}, which means that the maximum of the asymptotic ESS ratio of Laguerre-MPPI is no less than that of conventional MPPI.
Section~\ref{sec:exp3} verifies that the stability condition \eqref{eq:rho} derived in the infinite-sample limit, also predicts whether the closed-loop system under finite samples converges or diverges.
Section~\ref{sec:exp4} verifies the lower bound on the number of samples given in Corollary~\ref{cor:error}.
\tabref{tab:param} lists the parameters used in each experiment.
In all experiments, the reference coefficient and the covariance matrix in the coefficient space are set to $\tilde\varpi_{\Sel}=0$ and given by \eqref{eq:SigmaS}, respectively.

\begin{table}[t]
\centering
\caption{Parameters of the numerical experiments}
\label{tab:param}
\footnotesize\setlength{\tabcolsep}{3pt}
\begin{tabular}{c|ccc}
\hline
 & \ref{sec:exp1} & \ref{sec:exp2} & \ref{sec:exp3}, \ref{sec:exp4} \\
\hline
$M$ & 30 & 1000 & $10^5$ (\ref{sec:exp4}: 1009) \\
$T$ & 10 & 5--35 & 20 \\
$N$, $d$ & 3, 3 & 5, 5 & 5, 3 \\
$\Delta$ [s] & 0.1 & 0.1 & 0.1 \\
$a$ & 0.4 & 0.8 & 0.5 \\
$\lambda$ & 10 & 10 & 120 \\
$\Sigma_v$ & $4I_2$ & 4 & 4 \\
$W$ & $\mathrm{diag}(10,10,3,3)$ & $\mathrm{diag}(50,20)$ & $\mathrm{diag}(20,10)$ \\
$\Pi$ & $\mathrm{diag}(25,25,3,3)$ & $\mathrm{diag}(50,20)$ & $\mathrm{diag}(20,10)$ \\
Others & $\alpha=3\times10^5$ & $x_0=0$, $r=1$ & $x_0=[0.05\ 0.04]^\top$\\
& & & $\chi_{\mathrm{tol}}=10^{-2}$\\
\hline
\end{tabular}
\end{table}
\subsection{Obstacle Avoidance of a Mobile Robot}\label{sec:exp1}
Consider the trajectory generation of a mobile robot described by the two-dimensional double integrator
\begin{equation*}
x(k+1)=\begin{bmatrix}I_2&\Delta I_2\\0&I_2\end{bmatrix}x(k)+\begin{bmatrix}\frac{\Delta^2}{2}I_2\\\Delta I_2\end{bmatrix}u(k),
\end{equation*}
where $x=[p_x\ \ p_y\ \ \dot p_x\ \ \dot p_y]^\top$ is the state and $u=[u_x\ \ u_y]^\top$ is the input.
The initial and the target states are set to $x_0=[0\ \ 0\ \ 0\ \ 0]^\top$ and $x_{\mathrm{ref}}=[10\ \ 10\ \ 0\ \ 0]^\top$, respectively.
We let the state cost $S(V)$ be given by
\begin{align*}
S(V)=&\sum_{\tau=0}^{T-1}\bigl(\|x(\tau)-x_{\mathrm{ref}}\|_W^2+\alpha\,\mathbb{I}(x(\tau))\bigr)\\
&+\|x(T)-x_{\mathrm{ref}}\|_{\Pi}^2+\alpha\,\mathbb{I}(x(T)),
\end{align*}
where $\mathbb{I}(x)$ is the indicator function, i.e., $\mathbb{I}(x)=1$ if the position $(p_x,p_y)$ is inside any obstacle and $\mathbb{I}(x)=0$ otherwise.

\begin{figure}[t]
\centering
\subfloat[MPPI]{%
  \figorbox{0.7\linewidth}{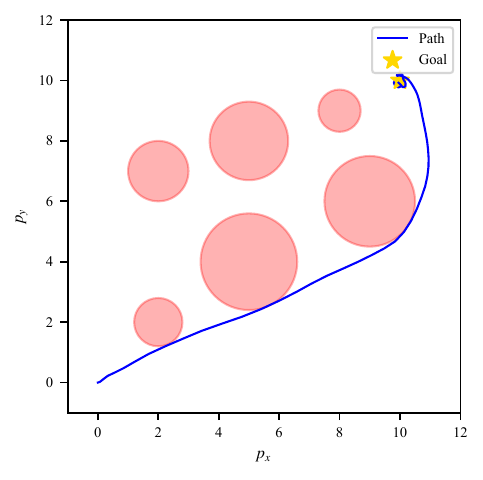}}\\
\subfloat[Laguerre-MPPI]{%
  \figorbox{0.7\linewidth}{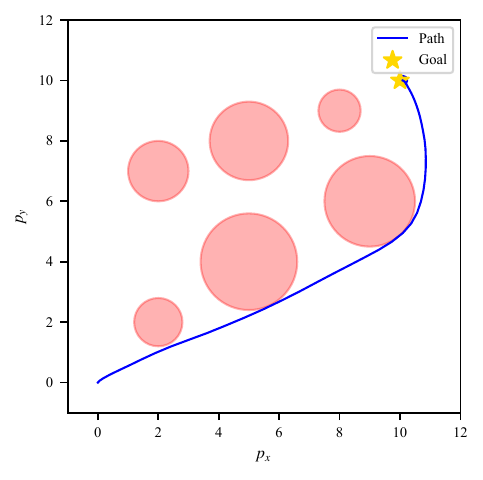}}
\caption{Trajectories in obstacle avoidance. Red circles represent obstacles.}\label{fig:traj}
\end{figure}

\figref{fig:traj} shows the resulting trajectories.
As illustrated in \figref{fig:traj}(a), the trajectory generated by conventional MPPI is jagged and circles around the target.
Since only 30 samples are used for the search dimension $mT=20$, high-frequency components due to the sampling error remain in the computed inputs.
As shown in \figref{fig:traj}(b), the trajectory generated by Laguerre-MPPI is smooth and shows no circling near the target.
Two factors cause this difference.
First, the search dimension is reduced to $md=6$, which makes 30 samples relatively dense.
Second, the basis functions with the pole $a>0$ have low-pass characteristics, which reduces the variation of the computed inputs.

To compare the smoothness discussed above quantitatively, we evaluate the inputs over the interval in which both methods follow nearly the same course.
The evaluation interval ends at the first time the robot comes within a distance of 0.5 of the target, and the relative input variation $\|u(k+1)-u(k)\|_2/\|u(k)\|_2$ is averaged over time and over 50 random seeds.
The arrival time is 3.93\,s for MPPI and 3.89\,s for Laguerre-MPPI, which are nearly equal. Over this interval, the relative variation is 1.354 for MPPI and 0.837 for Laguerre-MPPI, which is approximately 38\% smaller.
This result confirms that Laguerre-MPPI yields smoother inputs than conventional MPPI.
\subsection{Control Performance with Increasing Horizon}\label{sec:exp2}
We evaluate the step response of the one-dimensional double integrator
\begin{equation*}
x(k+1)=\begin{bmatrix}1&\Delta\\0&1\end{bmatrix}x(k)+\begin{bmatrix}\frac{\Delta^2}{2}\\\Delta\end{bmatrix}u(k),
\end{equation*}
where $x=[p\ \dot p]^\top$ represents the state consisting of the position $p$ and the velocity $\dot p$. The initial and the target states are set to $x_0=[0\ \ 0]^\top$ and $x_{\mathrm{ref}}=[r\ \ 0]^\top$, respectively, with the target position $r=1$.
The state cost is given by \eqref{eq:quad} with the state $x(k)$ replaced by the error $x(k)-x_{\mathrm{ref}}$.
We fix the number of samples $M=1000$ and the number of basis functions $N=5$, and vary the horizon $T$ from 5 to 35.
The control performance is evaluated by the index
\begin{equation*}
J_{\mathrm{IAE}}:=\sum_{k=0}^{K_{\mathrm{f}}}|r-p(k)|,
\end{equation*}
where $K_{\mathrm{f}}=50$.

\begin{figure}[t]
\centering
\figorbox{\linewidth}{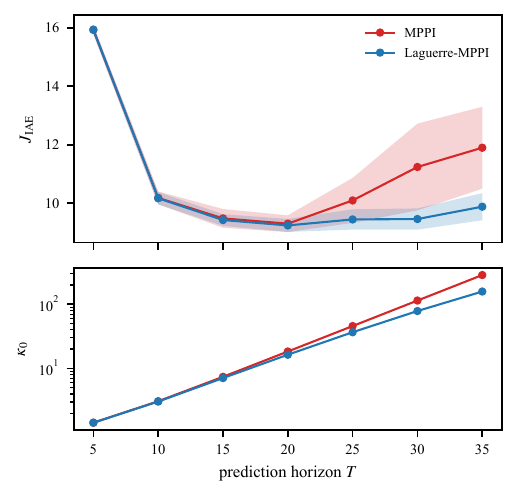}
\caption{Prediction horizon $T$ versus $J_{\mathrm{IAE}}$ (top, mean and $\pm1\sigma$ over 20 seeds) and versus $\kappa_0(5,I_5)$ of Laguerre-MPPI and $\kappa_0(T,I_{T})$ of conventional MPPI (bottom).}
\label{fig:sweep}
\end{figure}
The top panel of \figref{fig:sweep} shows the mean and the $\pm1\sigma$ range of the performance index $J_{\mathrm{IAE}}$ over 20 random seeds.
For horizons $T\le20$, MPPI and Laguerre-MPPI achieve comparable control performance, and the control performance of both methods improves as the horizon $T$ increases.
For horizons $T>20$, the control performance of MPPI degrades substantially, and its variation across seeds also grows.
In contrast, the control performance of Laguerre-MPPI degrades only slightly and varies little across seeds.
This difference arises from the search dimension reduction.
The search dimension of MPPI is $mT$ and grows linearly with $T$, whereas that of Laguerre-MPPI remains $md=5$ regardless of the horizon $T$.
Hence, Laguerre-MPPI is less susceptible to the curse of dimensionality caused by the extension of the horizon.

The bottom panel of \figref{fig:sweep} shows $\kappa_0(5,I_5)$ of Laguerre-MPPI and $\kappa_0(T,I_T)$, which corresponds to conventional MPPI as stated in Remark~\ref{rem:kappa0_mppi}, computed under the same setting.
As stated in 2) of Theorem~\ref{thm:kappa0}, $\kappa_0(5,I_5)\le\kappa_0(T,I_T)$ holds for all horizons, with equality at $T=5$.
\subsection{Basis Selection and Stability}\label{sec:exp3}
We verify the stability test based on Corollary~\ref{col:stab}. Consider the plant system given by the oscillator
\begin{equation*}
x(k+1)=\begin{bmatrix}\cos\Delta&\sin\Delta\\-\sin\Delta&\cos\Delta\end{bmatrix}x(k)+\begin{bmatrix}1-\cos\Delta\\\sin\Delta\end{bmatrix}u(k),
\end{equation*}
where $x=[x_1\ x_2]^\top=[p\ \dot p]^\top$ is the state, composed of the position $p$ and the velocity $\dot p$.
\tabref{tab:rho} shows the spectral radius $\rho(A-BK(\Sel))$ computed with the feedback gain \eqref{eq:K} for 10 selection matrices $\Sel$ that select $d=3$ basis functions out of $N=5$.
\begin{table}[t]
\centering
\caption{Basis selection and spectral radius.}
\label{tab:rho}
\begin{tabular}{cc|cc}
\hline
Basis set & $\rho(A-BK(\Sel))$ & Basis set & $\rho(A-BK(\Sel))$\\
\hline
$\{1,3,4\}$ & 0.904 & $\{1,2,4\}$ & 0.952\\
$\{1,4,5\}$ & 0.907 & $\{3,4,5\}$ & 0.989\\
$\{1,3,5\}$ & 0.915 & $\{2,3,5\}$ & \textbf{1.024}\\
$\{1,2,3\}$ & 0.943 & $\{2,3,4\}$ & \textbf{1.026}\\
$\{1,2,5\}$ & 0.949 & $\{2,4,5\}$ & \textbf{1.041}\\
\hline
\end{tabular}
\end{table}
Seven basis sets yield a spectral radius $\rho<1$, which implies closed-loop stability, while the three basis sets $\{2,3,4\}$, $\{2,3,5\}$, and $\{2,4,5\}$ yield a spectral radius $\rho>1$, which implies instability.
For example, for the set $\{2,3,4\}$, the eigenvalues of $A-BK(\Sel)$ in \eqref{eq:nominal} are $1.02\pm 0.099\mathrm{j}$, which predicts oscillatory divergence.

\begin{figure}[t]
\centering
\subfloat[Closed-loop behavior by Laguerre-MPPI with selected basis sets.]{%
  \figorbox{\linewidth}{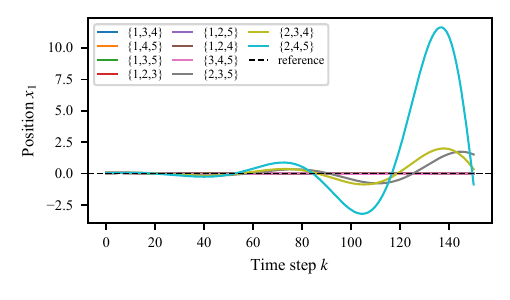}\label{fig:stabsim_all}}\\
\subfloat[Closed-loop behavior by Laguerre-MPPI with stabilizing basis sets.]{%
  \figorbox{\linewidth}{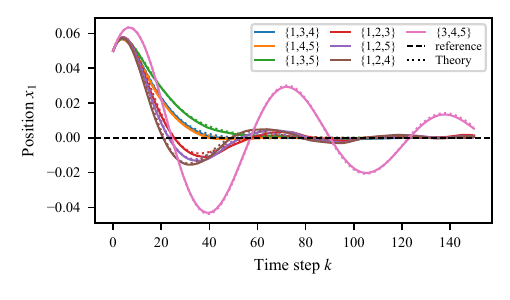}\label{fig:stabsim_stable}}
\caption{Position $x_1$ in the closed-loop responses of Laguerre-MPPI ($M=10^5$). In (b), dotted lines show the response of the infinite-sample limit \eqref{eq:nominal}.}
\label{fig:stabsim}
\end{figure}

\figref{fig:stabsim} shows the state trajectories generated by Laguerre-MPPI with the finite number of samples $M=10^5$.
In \figref{fig:stabsim}(a), the trajectories in the case that $\rho<1$ stay in a neighborhood of the origin, and the trajectories in the case that $\rho>1$ diverge with oscillation. The simulation results are consistent with the stability test summarized in \tabref{tab:rho}.
\figref{fig:stabsim}(b) shows an enlarged view of \figref{fig:stabsim}(a), allowing detailed inspection of the converging trajectories.
In the figure, the solid lines show representative trajectories generated by Laguerre-MPPI with a finite number of samples,
whereas the dotted lines show the theoretical trajectories computed from \eqref{eq:nominal}, corresponding to the infinite-sample limit.
The two trajectories agree closely in each case.
The error from the theoretical trajectory is caused by the stochastic noise $\varepsilon_M$, defined in \eqref{eq:eps}, and appears as small fluctuations near the origin.

\subsection{Number of Samples and Finite-Sample Behavior}\label{sec:exp4}
We verify the lower bound \eqref{eq:Mreq} in Corollary~\ref{cor:error} for the stabilizing basis set $\{1,3,4\}$ under the same setting as Section~\ref{sec:exp3}.
For the tolerance $\chi_{\mathrm{tol}}=10^{-2}$, the solution $\Upsilon$ of \eqref{eq:Lyap_eq}, computed from the design parameters, gives $M\ge1009$ in \eqref{eq:Mreq}.
We set the number of samples to the smallest value $M=1009$ satisfying \eqref{eq:Mreq} and run the closed-loop simulation for 500 random seeds.

\begin{figure}[t]
\centering
\figorbox{\linewidth}{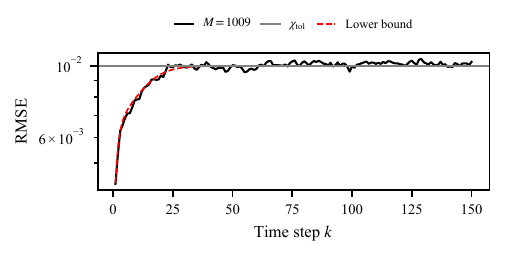}
\caption{RMSE $\E[\|\chi(k)\|^2]^{1/2}$ over 500 seeds for the basis set $\{1,3,4\}$ and $M=1009$ (solid), its lower bound $(\mathrm{tr}\,\Upsilon(k)/M)^{1/2}$ (dashed), and the tolerance $\chi_{\mathrm{tol}}$ (horizontal line).}
\label{fig:Mthreshold}
\end{figure}

\figref{fig:Mthreshold} shows the results.
The solid line is the RMSE $\E[\|\chi(k)\|^2]^{1/2}$ estimated over the 500 seeds.
The dashed line is $(\mathrm{tr}\,\Upsilon(k)/M)^{1/2}$, where $\Upsilon(k)$ is computed by iterating the right-hand side of \eqref{eq:Lyap_eq} from $\Upsilon(0)=0$.
As shown in the proof of Corollary~\ref{cor:error}, this curve is a lower bound on the RMSE at each time $k$ and converges to $(\mathrm{tr}\,\Upsilon/M)^{1/2}$ as $k\to\infty$.
The horizontal line shows the tolerance $\chi_{\mathrm{tol}}$.
As $k$ increases, the lower bound converges to $(\mathrm{tr}\,\Upsilon/M)^{1/2}$, which nearly equals $\chi_{\mathrm{tol}}$ for $M=1009$.
The measured RMSE is consistent with the lower bound over the whole time interval, and hence $(\mathrm{tr}\,\Upsilon/M)^{1/2}$ approximates the RMSE well in the steady state.

\section{CONCLUSIONS}\label{sec:conclusion}
We proposed Laguerre-MPPI, which expands the sampled inputs in Laguerre basis functions and samples the coefficients.
In addition to developing the algorithm, we derived a closed-loop stability condition in the infinite-sample limit for linear systems with quadratic costs and analyzed the sampling error and the ESS under finite samples.
Based on this analysis, we derived a lower bound on the number of samples for a prescribed deviation from the infinite-sample limit.
The obstacle avoidance experiment showed that Laguerre-MPPI generates smoother inputs than conventional MPPI. The comparison with MPPI in step responses showed that Laguerre-MPPI is robust to long prediction horizons.
We also showed that the closed-loop stability depends on the basis selection and verified the validity of the stability condition.

Future work includes a mean-square stability analysis of the stochastic system \eqref{eq:stoch}~\cite{Yoon2026} and a systematic basis-selection method that guarantees stability~\cite{3104482.3104615}.


\bibliographystyle{IEEEtran}
\bibliography{references}

\section*{Appendix}
\setcounter{section}{0}
\renewcommand{\thesection}{\Alph{section}}
\makeatletter
\def\thesectiondis{\Alph{section}.}
\makeatother

\section{Proof of Lemma~\ref{lemma:free_energy_bound}}\label{app:free_energy_bound}
\begin{proof}
\begin{align*}
    \mathcal{F}(S, \tilde{q},\lambda) &= -\lambda\ln \left(\mathbb{E}_{\tilde{\mathbb{Q}}} \left[ \exp\left(-\frac{S(V)}{\lambda}\right) \right]\right)\\
    &= -\lambda\ln \left(\mathbb{E}_{\mathbb{Q}} \left[ \exp\left(-\frac{S(V)}{\lambda}\right) \frac{\tilde{q}(V)}{q(V)} \right] \right).
\end{align*}
Jensen's inequality then gives
\begin{align}
    \mathcal{F}(S, \tilde{q},\lambda) &= -\lambda\ln \left(\mathbb{E}_{\mathbb{Q}} \left[ \exp\left(-\frac{S(V)}{\lambda}\right) \frac{\tilde{q}(V)}{q(V)} \right] \right)\notag\\
    &\leq -\lambda \mathbb{E}_{\mathbb{Q}} \left[ \ln \left( \exp\left(-\frac{S(V)}{\lambda}\right) \frac{\tilde{q}(V)}{q(V)} \right) \right].\label{eq:free_energy_bound_proof}
    \end{align}
The right-hand side of \eqref{eq:free_energy_bound_proof} satisfies
\begin{align*}
(\text{RHS}) &= -\lambda \mathbb{E}_{\mathbb{Q}} \left[ -\frac{S(V)}{\lambda} -\ln \left( \frac{q(V)}{\tilde{q}(V)} \right) \right]\\
    &= \mathbb{E}_{\mathbb{Q}}\left[S(V)\right] + \lambda \DKL(\mathbb{Q}\|\tilde{\mathbb{Q}}).
\end{align*}
Therefore \eqref{eq:free_energy_bound} holds.
\end{proof}

\section{Proof of Lemma~\ref{lemma:relaxed_SOC}}\label{app:relaxed_SOC}
\begin{proof}
Applying the control distribution $\mathbb{Q} = \mathcal{N}(U,\Sigma_{v})$ and the reference distribution $\tilde{\mathbb{Q}} = \mathcal{N}(\tilde{U},\Sigma_{v})$ to the KL divergence $\DKL(\mathbb{Q}\|\tilde{\mathbb{Q}})$ in \eqref{eq:J(Q)} yields
\begin{equation*}
  \begin{split}
&\DKL(\mathbb{Q}_U\|\tilde{\mathbb{Q}}) \\
=&\frac{1}{2}\sum_{k=0}^{T-1} \mathbb{E}_{\mathbb{Q}_U}\left[(v(k)-\tilde{u}(k))^\top\Sigma_{v}^{-1}(v(k)-\tilde{u}(k))\right.\\
&-\left.(v(k)-u(k))^\top\Sigma_{v}^{-1}(v(k)-u(k))\right].
  \end{split}
\end{equation*}
Using $\mathbb{E}_{\mathbb{Q}_U}[v(k)-u(k)]=0$, we obtain
\begin{equation*}
    \DKL(\mathbb{Q}_U\|\tilde{\mathbb{Q}}) = \frac{1}{2}\sum_{k=0}^{T-1} (u(k)-\tilde{u}(k))^\top\Sigma_{v}^{-1}(u(k)-\tilde{u}(k)).
\end{equation*}
Hence, under Assumptions~\ref{assumption:cost_decomposition} and \ref{assumption:quadratic_input_cost},
\begin{equation*}
  \begin{split}
    &\mathbb{E}_{\mathbb{Q}_U}\left[S(V)\right] + \lambda \DKL(\mathbb{Q}_U\|\tilde{\mathbb{Q}}) \\
    &= \mathbb{E}_{\mathbb{Q}_U}\left[\phi(x(T))+\sum_{k=0}^{T-1} \mathcal{L}_k(x(k), u(k))\right].
  \end{split}
\end{equation*}
\end{proof}

\section{Proof of Lemma~\ref{lemma:optimal_distribution}}\label{app:optimal_distribution}
\begin{proof}
We transform the objective function of the relaxed problem for a distribution $\mathbb{Q}$ of arbitrary shape.
Since
$\exp\left(-\frac{S(V)}{\lambda}\right)\tilde{q}(V)=\eta\, q^*(V)$ by \eqref{eq:optimal_control_distribution},
\begin{align*}
  &\mathbb{E}_{\mathbb{Q}}\left[S(V)\right] + \lambda \DKL(\mathbb{Q}\|\tilde{\mathbb{Q}})\\
  &= \lambda\int_{\mathbb{R}^{m\times T}} q(V)\ln\left(\frac{q(V)}{\exp\left(-\frac{S(V)}{\lambda}\right)\tilde{q}(V)}\right) dV \\
  &= \lambda\int_{\mathbb{R}^{m\times T}} q(V)\ln\left(\frac{q(V)}{q^*(V)}\right) dV - \lambda\ln\eta \\
  &= \lambda\DKL(\mathbb{Q}\|\mathbb{Q}^*) + \mathcal{F}(S,\tilde{q},\lambda).
\end{align*}
The last equality follows from the definition \eqref{eq:free_energy} of the free energy and \eqref{eq:eta}. Next, by the two properties of the KL divergence
\begin{itemize}
    \item  $\DKL(\mathbb{F}\|\mathbb{H})\geq 0$,
    \item  $\DKL(\mathbb{F}\|\mathbb{H})=0 \Leftrightarrow \mathbb{F}=\mathbb{H}$,
\end{itemize}
the KL divergence satisfies $\DKL(\mathbb{Q}\|\mathbb{Q}^*)=0$ only when the distribution equals the optimal distribution, $\mathbb{Q}=\mathbb{Q}^*$.
Therefore the cost function $J(\mathbb{Q})$ attains its minimum value, the free energy $\mathcal{F}(S,\tilde{q},\lambda)$, if and only if $\mathbb{Q}=\mathbb{Q}^*$.
\end{proof}

\section{Proof of Theorem~\ref{theorem:MPPI_importance_sampling}}\label{app:MPPI_importance_sampling}
\begin{proof}
Let
\begin{equation}\label{eq:wtilde}
\tilde{w}:=\exp\left(-\frac{S(V)}{\lambda}\right)\frac{\tilde{q}(V)}{\hat{q}(V)}.
\end{equation}
Then the density of the optimal distribution is $q^*(V)=\frac{\tilde w\,\hat{q}(V)}{\eta}$. Therefore
\begin{equation*}
  \mathbb{E}_{\mathbb{Q}^*}[V] = \frac{1}{\eta}\mathbb{E}_{\hat{\mathbb{Q}}}[\tilde w\,V],\quad
  \eta = \mathbb{E}_{\hat{\mathbb{Q}}}[\tilde w].
\end{equation*}
Applying the initial estimate distribution $\hat{\mathbb{Q}}=\mathcal{N}(\hat{U},\Sigma_v)$ and the reference distribution $\tilde{\mathbb{Q}}=\mathcal{N}(\tilde{U},\Sigma_v)$, $\tilde w$ defined in \eqref{eq:wtilde} factorizes as
\begin{align*}
\tilde w(V) &= \exp\left(-\frac12\sum_{k}\mathcal{D}_k\right) w(V),\\
\mathcal{D}_k &:= \tilde{u}(k)^\top\Sigma_v^{-1}\tilde{u}(k) - \hat{u}(k)^\top\Sigma_v^{-1}\hat{u}(k).
\end{align*}
Since the factor $\exp\left(-\frac12\sum_k\mathcal D_k\right)$ does not depend on the sampled input sequence $V$,
\begin{equation*}
  \mathbb{E}_{\mathbb{Q}^*}[V] = \frac{\mathbb{E}_{\hat{\mathbb{Q}}}[\tilde w\,V]}{\mathbb{E}_{\hat{\mathbb{Q}}}[\tilde w]}
  = \frac{\mathbb{E}_{\hat{\mathbb{Q}}}[w\,V]}{\mathbb{E}_{\hat{\mathbb{Q}}}[w]}.
\end{equation*}
\end{proof}

\section{Proof of Lemma~\ref{lem:moment}}\label{app:proof_lem_moment}
\begin{proof}
  \textit{1)}
Since $\Sigma_{\Sel}^{-1}\succ0$, $H_{\Sel}\succeq0$, and $p>0$, \eqref{eq:Lambda_p} gives $\Lambda_p\succ0$.
The outline of the proof is as follows. Since both the $p$-th power $w_{\Sel}^p$ of the weight and the density $\hat{q}_{\Sel}$ are exponentials of quadratic functions of the selected coefficient $\vartheta_{\Sel}$,
their product $w_{\Sel}^p\hat{q}_{\Sel}$ is written as a constant multiple of a Gaussian probability density, which reduces $Z_p=\mathbb{E}_{\hat{\mathbb{Q}}_{\Sel}}[w_{\Sel}^p]$ defined in \eqref{eq:Zdef} to an expectation under $\mathbb{Q}_p$ defined in \eqref{eq:Qp}.

We rearrange the integrand $w_{\Sel}^p\hat{q}_{\Sel}$.
By \eqref{eq:Lw} and \eqref{eq:Squad}, the $p$-th power $w_{\Sel}^p$ of the weight is
\begin{align}
w_{\Sel}^p(\vartheta_{\Sel})=\exp\Bigl(&-\frac{p}{\lambda}\bigl(\|\vartheta_{\Sel}\|^2_{H_{\Sel}}+2(G_{\Sel}x_0)^\top\vartheta_{\Sel}\bigr)\notag\\
&-p(\hat\varpi_{\Sel}-\tilde\varpi_{\Sel})^\top\Sigma_{\Sel}^{-1}\vartheta_{\Sel}\Bigr).\label{eq:wp}
\end{align}
The probability density of the initial estimate distribution $\hat{\mathbb{Q}}_{\Sel}$ is
\begin{equation}\label{eq:qhat}
\hat{q}_{\Sel}(\vartheta_{\Sel})=c\exp\Bigl(-\frac12\|\vartheta_{\Sel}-\hat\varpi_{\Sel}\|_{\Sigma_{\Sel}^{-1}}^2\Bigr),
\end{equation}
where
\begin{equation}\label{eq:c}
c:=(2\pi)^{-md/2}|\Sigma_{\Sel}|^{-1/2}.
\end{equation}
Multiplying \eqref{eq:wp} by \eqref{eq:qhat} and rearranging in the selected coefficient $\vartheta_{\Sel}$ with \eqref{eq:Lambda_p} and \eqref{eq:b_p} yields
\begin{equation}\label{eq:wq}
w_{\Sel}^p(\vartheta_{\Sel})\hat{q}_{\Sel}(\vartheta_{\Sel})=c\exp\Bigl(-\frac12\|\vartheta_{\Sel}\|_{\Lambda_p}^2-b_p^\top\vartheta_{\Sel}-\frac12\|\hat\varpi_{\Sel}\|_{\Sigma_{\Sel}^{-1}}^2\Bigr).
\end{equation}
Completing the square in the exponent of \eqref{eq:wq} with $m_p$ defined in \eqref{eq:m_p} gives
\begin{align*}
&-\frac12\|\vartheta_{\Sel}\|_{\Lambda_p}^2-b_p^\top\vartheta_{\Sel}-\frac12\|\hat\varpi_{\Sel}\|_{\Sigma_{\Sel}^{-1}}^2\\
&=-\frac12\|\vartheta_{\Sel}-m_p\|_{\Lambda_p}^2+\frac12\|b_p\|_{\Lambda_p^{-1}}^2-\frac12\|\hat\varpi_{\Sel}\|_{\Sigma_{\Sel}^{-1}}^2.
\end{align*}
Therefore
\begin{align}
w_{\Sel}^p(\vartheta_{\Sel})\,\hat{q}_{\Sel}(\vartheta_{\Sel})&=\left\{c\exp\Bigl(\frac12\|b_p\|_{\Lambda_p^{-1}}^2-\frac12\|\hat\varpi_{\Sel}\|_{\Sigma_{\Sel}^{-1}}^2\Bigr)\right\}\notag\\
&\cdot \exp\Bigl(-\frac12\|\vartheta_{\Sel}-m_p\|_{\Lambda_p}^2\Bigr).\label{eq:wq_gauss}
\end{align}
We now integrate \eqref{eq:wq_gauss} over $\mathbb{R}^{md}$ using the Gaussian integral
\begin{equation}\label{eq:gauss_integral}
\int_{\mathbb{R}^{md}}\exp\Bigl(-\frac12\|\vartheta_{\Sel}-m_p\|_{\Lambda_p}^2\Bigr)\,d\vartheta_{\Sel}=(2\pi)^{\frac{md}{2}}|\Lambda_p|^{-\frac{1}{2}}.
\end{equation}
By the definition \eqref{eq:Zdef} of $Z_p$, \eqref{eq:wq_gauss}, and \eqref{eq:gauss_integral},
\begin{align*}
Z_p&=\int_{\mathbb{R}^{md}}w_{\Sel}^p(\vartheta_{\Sel})\hat{q}_{\Sel}(\vartheta_{\Sel})\,d\vartheta_{\Sel}\\
&=c\,(2\pi)^{\frac{md}{2}}|\Lambda_p|^{-\frac{1}{2}}\exp\Bigl(\frac12\|b_p\|_{\Lambda_p^{-1}}^2-\frac12\|\hat\varpi_{\Sel}\|_{\Sigma_{\Sel}^{-1}}^2\Bigr).
\end{align*}
Substituting $c$ defined in \eqref{eq:c} yields \eqref{eq:Zp}. The right-hand side of \eqref{eq:Zp} is positive and finite.

\textit{2)}
Since the probability density of $\mathbb{Q}_p$ defined in \eqref{eq:Qp} is
\begin{equation}\label{eq:qp}
q_p(\vartheta_{\Sel})=(2\pi)^{-\frac{md}{2}}|\Lambda_p|^{-\frac{1}{2}}\exp\Bigl(-\frac12\|\vartheta_{\Sel}-m_p\|_{\Lambda_p}^2\Bigr),
\end{equation}
expressing the right-hand side of \eqref{eq:wq_gauss} obtained in the proof of 1) with the closed form \eqref{eq:Zp} of $Z_p$ and \eqref{eq:qp} yields \eqref{eq:wq_density}.

For $p=1$, the density ratio in the proof of 4) of Theorem~\ref{thm:coefMPPI} shows that the product $w_{\Sel}\hat{q}_{\Sel}$ equals $\exp(-S_{\vartheta_{\Sel}}/\lambda)\,\tilde q_{\Sel}$ up to a positive constant factor independent of the selected coefficient $\vartheta_{\Sel}$.
The redefinition of $S_{\vartheta_{\Sel}}$ in \eqref{eq:Squad} removes only a constant term and hence does not change the density $q_{\Sel}^*$ of the optimal distribution given by \eqref{eq:qstar}.
Hence, by \eqref{eq:qstar}, the product $w_{\Sel}\hat{q}_{\Sel}$ is proportional to the density $q_{\Sel}^*$.
By \eqref{eq:wq_density} with $p=1$, the product $w_{\Sel}\hat{q}_{\Sel}$ is also proportional to the density $q_1$.
Since both $q_{\Sel}^*$ and $q_1$ are probability densities, $q_1=q_{\Sel}^*$, that is, $\mathbb{Q}_1=\mathbb{Q}_{\Sel}^*$.

\textit{3)}
By \eqref{eq:wq_density}, for any selected coefficient $\vartheta_{\Sel}$,
\begin{equation}\label{eq:abs_integrand}
|w_{\Sel}^p(\vartheta_{\Sel})\varphi(\vartheta_{\Sel})|\,\hat{q}_{\Sel}(\vartheta_{\Sel})=Z_p\,|\varphi(\vartheta_{\Sel})|\,q_p(\vartheta_{\Sel}).
\end{equation}
By the assumption $\E_{\mathbb{Q}_p}[|\varphi(\vartheta_{\Sel})|]<\infty$ and \eqref{eq:abs_integrand},
\begin{equation*}
\mathbb{E}_{\hat{\mathbb{Q}}_{\Sel}}[|w_{\Sel}^p(\vartheta_{\Sel})\varphi(\vartheta_{\Sel})|]<\infty,\quad
\mathbb{E}_{\mathbb{Q}_p}[|\varphi(\vartheta_{\Sel})|]<\infty.
\end{equation*}
Hence both sides of \eqref{eq:change_measure} are finite, and by \eqref{eq:wq_density},
\begin{align*}
\E_{\hat{\mathbb{Q}}_{\Sel}}[w_{\Sel}^p\varphi(\vartheta_{\Sel})]
&=\int_{\mathbb{R}^{md}}w_{\Sel}^p(\vartheta_{\Sel})\,\varphi(\vartheta_{\Sel})\,\hat{q}_{\Sel}(\vartheta_{\Sel})\,d\vartheta_{\Sel}\\
&=Z_p\int_{\mathbb{R}^{md}}\varphi(\vartheta_{\Sel})\,q_p(\vartheta_{\Sel})\,d\vartheta_{\Sel}\\
&=Z_p\,\E_{\mathbb{Q}_p}[\varphi(\vartheta_{\Sel})].
\end{align*}

\textit{4)}
The first and second moments of the Gaussian distribution $\mathbb{Q}_p$ are
\begin{equation}\label{eq:gauss_moment}
\E_{\mathbb{Q}_p}[\vartheta_{\Sel}]=m_p,\quad
\E_{\mathbb{Q}_p}[\vartheta_{\Sel}\vartheta_{\Sel}^\top]=\Lambda_p^{-1}+m_pm_p^\top,
\end{equation}
both of which are finite. Applying \eqref{eq:change_measure} to $\varphi(\vartheta_{\Sel})=\vartheta_{\Sel}$ and to each entry of $\vartheta_{\Sel}\vartheta_{\Sel}^\top$ and substituting \eqref{eq:gauss_moment} yield \eqref{eq:moment1} and \eqref{eq:moment2}.

\textit{5)}
Since all moments of a Gaussian distribution are finite, $\E_{\mathbb{Q}_p}[\|\vartheta_{\Sel}\|^r]<\infty$ for any $r\ge0$. Applying \eqref{eq:change_measure} to $\varphi(\vartheta_{\Sel})=\|\vartheta_{\Sel}\|^r$ gives
\begin{equation*}
\E_{\hat{\mathbb{Q}}_{\Sel}}[w_{\Sel}^p\|\vartheta_{\Sel}\|^r]=Z_p\,\E_{\mathbb{Q}_p}[\|\vartheta_{\Sel}\|^r]<\infty,
\end{equation*}
which yields \eqref{eq:moment_r}.
\end{proof}

\section{Proof of Theorem~\ref{thm:clt}}\label{app:proof_thm_clt}
\begin{proof}
Let the samples $\vartheta_{\Sel}^1,\ldots,\vartheta_{\Sel}^M$ be generated independently from the initial estimate distribution $\hat{\mathbb{Q}}_{\Sel}$, and recall from \eqref{eq:ND} that the finite-sample approximation $\varpi_{\Sel}^\dagger$ is the ratio $\bar N_M/\bar D_M$.

\textit{1)}
The outline is to apply the central limit theorem to the pair of the numerator $\bar N_M$ and the denominator $\bar D_M$, and to transfer the result to the convergence in distribution of the ratio $\bar N_M/\bar D_M$ by the delta method~\cite{vanderVaart1998}.

Recall from \eqref{eq:moment_r} that, $\E_{\hat{\mathbb{Q}}_{\Sel}}[w_{\Sel}^2]<\infty$ and $\E_{\hat{\mathbb{Q}}_{\Sel}}[w_{\Sel}^2\|\vartheta_{\Sel}\|^2]<\infty$.
Hence the independent and identically distributed random vectors
\begin{equation*}
[w_{\Sel}(\vartheta_{\Sel}^i)\vartheta_{\Sel}^{i\top}\ \ w_{\Sel}(\vartheta_{\Sel}^i)]^\top
\end{equation*}
have a finite covariance matrix.
By the definition \eqref{eq:Zdef} of $Z_1$ and \eqref{eq:moment1}, their mean is $[Z_1m_1^\top\ \ Z_1]^\top$.
The multivariate central limit theorem gives
\begin{equation}\label{eq:clt_ND}
  \begin{split}
\sqrt{M}&\left(\begin{bmatrix}\bar N_M\\\bar D_M\end{bmatrix}-\begin{bmatrix}Z_1m_1\\Z_1\end{bmatrix}\right)\\
&\xrightarrow{d}\mathcal{N}\left(0, \mathrm{Cov}_{\hat{\mathbb{Q}}_{\Sel}}\left[\begin{bmatrix}w_{\Sel}\vartheta_{\Sel}\\w_{\Sel}\end{bmatrix}\right]\right).
  \end{split}
\end{equation}
Applying the delta method to this result, we derive the covariance of the ratio $\bar{N}_M/\bar D_M$.

The ratio map $(\bar N_M,\bar D_M)\mapsto\bar N_M/\bar D_M$ maps the $(md+1)$-dimensional vector formed by the $md$-dimensional vector $\bar N_M$ and the positive scalar $\bar D_M$ to an $md$-dimensional vector.
Hence its Jacobian matrix is an $md\times(md+1)$ matrix that consists of the partial derivatives with respect to the sample means $\bar N_M$ and $\bar D_M$ placed side by side.
The partial derivative of the ratio $\bar N_M/\bar D_M$ with respect to the numerator $\bar N_M$ is
\begin{equation}\label{eq:jacobian_N}
\frac{\partial}{\partial\bar N_M}\left[\frac{\bar N_M}{\bar D_M}\right]=\frac{1}{\bar D_M}I_{md},
\end{equation}
and the partial derivative of the ratio $\bar N_M/\bar D_M$ with respect to the denominator $\bar D_M$ is
\begin{equation}\label{eq:jacobian_D}
\frac{\partial}{\partial\bar D_M}\left[\frac{\bar N_M}{\bar D_M}\right]=-\frac{\bar N_M}{\bar D_M^2}.
\end{equation}
Placing \eqref{eq:jacobian_N} and \eqref{eq:jacobian_D} side by side and evaluating them at the point $(Z_1m_1,Z_1)$, we obtain the Jacobian matrix
\begin{equation}\label{eq:jacobian}
\mathcal{J}=\left[\frac{1}{Z_1}I_{md}\ \ -\frac{Z_1m_1}{Z_1^2}\right]=\frac{1}{Z_1}\bigl[I_{md}\ \ -m_1\bigr].
\end{equation}
Since $Z_1>0$, the ratio map is differentiable at the point $(Z_1m_1,Z_1)$, and its value at this point is the optimal coefficient, $Z_1m_1/Z_1=m_1=\varpi_{\Sel}^*$.

Applying the delta method~\cite{vanderVaart1998} to \eqref{eq:clt_ND} yields
\begin{equation}\label{eq:delta}
\sqrt{M}(\varpi_{\Sel}^\dagger-\varpi_{\Sel}^*)\xrightarrow{d}\mathcal{N}\left(0,\ \mathrm{Cov}_{\hat{\mathbb{Q}}_{\Sel}}\left[\mathcal{J}\begin{bmatrix}w_{\Sel}\vartheta_{\Sel}\\w_{\Sel}\end{bmatrix}\right]\right).
\end{equation}
Using \eqref{eq:jacobian}, we have
\begin{equation}\label{eq:J_xi}
\mathcal{J}\begin{bmatrix}w_{\Sel}(\vartheta_{\Sel})\vartheta_{\Sel}\\w_{\Sel}(\vartheta_{\Sel})\end{bmatrix}=\frac{w_{\Sel}(\vartheta_{\Sel})}{Z_1}(\vartheta_{\Sel}-m_1).
\end{equation}
From the definition \eqref{eq:Zdef} of $Z_1$ and \eqref{eq:moment1}, we see that
\begin{equation*}
\E_{\hat{\mathbb{Q}}_{\Sel}}\left[\frac{w_{\Sel}}{Z_1}(\vartheta_{\Sel}-m_1)\right]=\frac{1}{Z_1}(Z_1m_1-Z_1m_1)=0
\end{equation*}
holds.
Hence the covariance matrix of \eqref{eq:J_xi} equals its second moment, that is,
\begin{equation}\label{eq:Xi_expect}
  \begin{split}
&\mathrm{Cov}_{\hat{\mathbb{Q}}_{\Sel}}\left[\frac{w_{\Sel}}{Z_1}(\vartheta_{\Sel}-m_1)\right]\\
&=\frac{1}{Z_1^2}\E_{\hat{\mathbb{Q}}_{\Sel}}\bigl[w_{\Sel}^2(\vartheta_{\Sel}-m_1)(\vartheta_{\Sel}-m_1)^\top\bigr].
  \end{split}
\end{equation}
Expanding the expectation on the right-hand side of \eqref{eq:Xi_expect} and applying the definition \eqref{eq:Zdef} of $Z_2$, \eqref{eq:moment1}, and \eqref{eq:moment2} yields
\begin{align*}
&\E_{\hat{\mathbb{Q}}_{\Sel}}\bigl[w_{\Sel}^2(\vartheta_{\Sel}-m_1)(\vartheta_{\Sel}-m_1)^\top\bigr]\\
&=Z_2\bigl[\Lambda_2^{-1}+m_2m_2^\top-m_2m_1^\top-m_1m_2^\top+m_1m_1^\top\bigr]\\
&=Z_2\bigl[\Lambda_2^{-1}+(m_2-m_1)(m_2-m_1)^\top\bigr].
\end{align*}
Substituting this into \eqref{eq:Xi_expect} and using $\kappa^{\Sel}=Z_2/Z_1^2$, defined in \eqref{eq:kappaL}, the covariance matrix in \eqref{eq:delta} is reduced to $\Xi(\kappa^{\Sel})$ of \eqref{eq:Xi}.

\textit{2)}
From \eqref{eq:ESS}, we have
\begin{equation}\label{eq:ESS_ratio}
\frac{\ESS}{M}=\frac{\bigl[\frac1M\sum_{i=1}^Mw_{\Sel}(\vartheta_{\Sel}^i)\bigr]^2}{\frac1M\sum_{i=1}^Mw_{\Sel}(\vartheta_{\Sel}^i)^2}.
\end{equation}
It follows from \eqref{eq:Zp} that both $\E_{\hat{\mathbb{Q}}_{\Sel}}[w_{\Sel}]=Z_1$ and $\E_{\hat{\mathbb{Q}}_{\Sel}}[w_{\Sel}^2]=Z_2$ are positive and finite.
By the strong law of large numbers, the numerator and the denominator of the right-hand side of \eqref{eq:ESS_ratio} converge almost surely to $Z_1^2$ and $Z_2$, respectively.
This shows that $\ESS/M$ converges almost surely to $Z_1^2/Z_2=1/\kappa^{\Sel}$.

\textit{3)}
The outline is to substitute the closed form \eqref{eq:Zp} of $Z_p$ in Lemma~\ref{lem:moment} into $\kappa^{\Sel}=Z_2/Z_1^2$, defined in \eqref{eq:kappaL}, and to express the exponent in terms of $\nu$, defined in \eqref{eq:e}.
Applying \eqref{eq:Zp} gives
\begin{equation}\label{eq:kappa_Z}
\kappa^{\Sel}=\frac{Z_2}{Z_1^2}=\frac{|\Sigma_{\Sel}|^{1/2}|\Lambda_1|}{|\Lambda_2|^{1/2}}\exp\left(E\right),
\end{equation}
where the exponent $E$ is
\begin{equation}\label{eq:kappa_E}
E:=\frac12\|b_2\|_{\Lambda_2^{-1}}^2-\|b_1\|_{\Lambda_1^{-1}}^2+\frac12\|\hat\varpi_{\Sel}\|_{\Sigma_{\Sel}^{-1}}^2.
\end{equation}

We simplify the coefficient in front of the exponential function in \eqref{eq:kappa_Z}.
By using \eqref{eq:Lambda_p}, the determinant of $\Lambda_p$ is $|\Lambda_p|=|\Sigma_{\Sel}|^{-1}\bigl|I_{md}+\frac{2p}{\lambda}\Sigma_{\Sel}H_{\Sel}\bigr|$.
Substituting this into \eqref{eq:kappa_Z} yields \eqref{eq:kappa0}.

We express the exponent \eqref{eq:kappa_E} in terms of $\nu$.
Using the definition \eqref{eq:m_p} of $m_p$, $b_1$ in \eqref{eq:b_p} satisfies $b_1=-\Lambda_1m_1$.
Furthermore, $b_2$ satisfies
\begin{equation*}
b_2=2b_1+\Sigma_{\Sel}^{-1}\hat\varpi_{\Sel}.
\end{equation*}
Using \eqref{eq:LIS} and \eqref{eq:moment1} with $p=1$, the optimal coefficient satisfies $\varpi_{\Sel}^*=m_1$.
Hence, using \eqref{eq:e}, $\hat\varpi_{\Sel}=m_1+\nu$ holds.
Using this relation and the identity
\begin{equation*}
2\Lambda_1-\Sigma_{\Sel}^{-1}=\Lambda_2
\end{equation*}
obtained from \eqref{eq:Lambda_p}, we have
\begin{equation}\label{eq:b2_e}
b_2=-2\Lambda_1m_1+\Sigma_{\Sel}^{-1}(m_1+\nu)=-\Lambda_2m_1+\Sigma_{\Sel}^{-1}\nu.
\end{equation}
Substituting $b_1=-\Lambda_1m_1$, \eqref{eq:b2_e}, and $\hat\varpi_{\Sel}=m_1+\nu$ into each term of \eqref{eq:kappa_E} gives
\begin{align*}
\frac12\|b_2\|_{\Lambda_2^{-1}}^2&=\frac12m_1^\top\Lambda_2m_1-m_1^\top\Sigma_{\Sel}^{-1}\nu+\frac12\nu^\top\Sigma_{\Sel}^{-1}\Lambda_2^{-1}\Sigma_{\Sel}^{-1}\nu,\\
\|b_1\|_{\Lambda_1^{-1}}^2&=m_1^\top\Lambda_1m_1,\\
\frac12\|\hat\varpi_{\Sel}\|_{\Sigma_{\Sel}^{-1}}^2&=\frac12m_1^\top\Sigma_{\Sel}^{-1}m_1+m_1^\top\Sigma_{\Sel}^{-1}\nu+\frac12\nu^\top\Sigma_{\Sel}^{-1}\nu.
\end{align*}
Therefore, we have the expression
\begin{equation*}
E=\frac12\nu^\top\bigl[\Sigma_{\Sel}^{-1}+\Sigma_{\Sel}^{-1}\Lambda_2^{-1}\Sigma_{\Sel}^{-1}\bigr]\nu=\frac12\nu^\top\Psi \nu.
\end{equation*}
This shows that \eqref{eq:kappa} holds.
Moreover,
since $\Sigma_{\Sel}^{-1}\succ0$ and $\Sigma_{\Sel}^{-1}\Lambda_2^{-1}\Sigma_{\Sel}^{-1}\succeq0$, $\Psi$ defined in \eqref{eq:Psi} is positive definite.
\end{proof}

\section{Details on $\kappa_0$}\label{app:details_kappa0}
This section relates $\kappa_0^{\Sel}$, defined in \eqref{eq:kappa0}, to the difference between the covariance matrix $\Sigma_{\Sel}$ of the initial estimate distribution $\hat{\mathbb{Q}}_{\Sel}$ and the covariance matrix $\Lambda_1^{-1}$ of the optimal distribution $\mathbb{Q}_{\Sel}^*$.
Let $\Sigma_{\Sel}^{1/2}$ be the symmetric square root of $\Sigma_{\Sel}$.
Since the matrix $\Sigma_{\Sel}^{1/2}H_{\Sel}\Sigma_{\Sel}^{1/2}$ is symmetric positive semidefinite, it is diagonalized by an orthogonal matrix $O$ as
\begin{equation*}
\Sigma_{\Sel}^{1/2}H_{\Sel}\Sigma_{\Sel}^{1/2}=O\,\mathrm{diag}\bigl(\mu_1^{\Sel},\ldots,\mu_{md}^{\Sel}\bigr)\,O^\top,
\end{equation*}
where $\mu_i^{\Sel}\ge0$ are its eigenvalues.
Since $\Sigma_{\Sel}H_{\Sel}=\Sigma_{\Sel}^{1/2}\bigl(\Sigma_{\Sel}^{1/2}H_{\Sel}\Sigma_{\Sel}^{1/2}\bigr)\Sigma_{\Sel}^{-1/2}$, these are also the eigenvalues of $\Sigma_{\Sel}H_{\Sel}$.

Using \eqref{eq:Lambda_p} and $I_{md}=OO^\top$, we factorize $\Lambda_1$ as
\begin{align*}
\Lambda_1&=\Sigma_{\Sel}^{-1/2}\Bigl(I_{md}+\frac{2}{\lambda}\Sigma_{\Sel}^{1/2}H_{\Sel}\Sigma_{\Sel}^{1/2}\Bigr)\Sigma_{\Sel}^{-1/2}\\
&=\Sigma_{\Sel}^{-1/2}O\Bigl(I_{md}+\frac{2}{\lambda}\mathrm{diag}\bigl(\mu_1^{\Sel},\ldots,\mu_{md}^{\Sel}\bigr)\Bigr)O^\top\Sigma_{\Sel}^{-1/2}.
\end{align*}
Then, we have $\Lambda_1^{-1}$ as
\begin{equation*}
\Lambda_1^{-1}=\Sigma_{\Sel}^{1/2}O\Bigl(I_{md}+\frac{2}{\lambda}\mathrm{diag}\bigl(\mu_1^{\Sel},\ldots,\mu_{md}^{\Sel}\bigr)\Bigr)^{-1}O^\top\Sigma_{\Sel}^{1/2}.
\end{equation*}
Consider the transformed coefficient $O^\top\Sigma_{\Sel}^{-1/2}\vartheta_{\Sel}$.
Its covariance matrices under $\hat{\mathbb{Q}}_{\Sel}$ and $\mathbb{Q}_{\Sel}^*$ are
\begin{align*}
&O^\top\Sigma_{\Sel}^{-1/2}\,\Sigma_{\Sel}\,\Sigma_{\Sel}^{-1/2}O=I_{md},\\
&O^\top\Sigma_{\Sel}^{-1/2}\Lambda_1^{-1}\Sigma_{\Sel}^{-1/2}O\\
&\quad=\mathrm{diag}\Bigl(\frac{1}{1+2\mu_1^{\Sel}/\lambda},\ldots,\frac{1}{1+2\mu_{md}^{\Sel}/\lambda}\Bigr),
\end{align*}
respectively.
Thus both covariance matrices are diagonal in these coordinates, and the ratio of the variance under $\hat{\mathbb{Q}}_{\Sel}$ to that under $\mathbb{Q}_{\Sel}^*$ in the $i$-th coordinate is $1+2\mu_i^{\Sel}/\lambda\ge1$.
Subtracting the two diagonal forms gives
\begin{align*}
&O^\top\Sigma_{\Sel}^{-1/2}\bigl(\Sigma_{\Sel}-\Lambda_1^{-1}\bigr)\Sigma_{\Sel}^{-1/2}O\\
&\quad=\mathrm{diag}\Bigl(\frac{2\mu_1^{\Sel}/\lambda}{1+2\mu_1^{\Sel}/\lambda},\ldots,\frac{2\mu_{md}^{\Sel}/\lambda}{1+2\mu_{md}^{\Sel}/\lambda}\Bigr).
\end{align*}
Therefore, $\Sigma_{\Sel}\succeq\Lambda_1^{-1}$ always holds, $\Sigma_{\Sel}\succ\Lambda_1^{-1}$ holds if and only if $\mu_i^{\Sel}>0$ for all $i$, and $\Sigma_{\Sel}=\Lambda_1^{-1}$ holds if and only if $\mu_i^{\Sel}=0$ for all $i$.

Since the eigenvalues of $I_{md}+c\Sigma_{\Sel}H_{\Sel}$ are $1+c\mu_i^{\Sel}$ for $c>0$, \eqref{eq:kappa0} gives
$\kappa_0^{\Sel}=\prod_{i=1}^{md}\psi(\mu_i^{\Sel})$ with the function $\psi$ defined in \eqref{eq:psi}.
For $\mu\ge0$, $(1+2\mu/\lambda)^2=1+4\mu/\lambda+4\mu^2/\lambda^2$ gives $\psi(\mu)\ge1$, with equality if and only if $\mu=0$.
Therefore $\kappa_0^{\Sel}=1$ holds if and only if $\mu_i^{\Sel}=0$ for all $i$, that is, $\Sigma_{\Sel}=\Lambda_1^{-1}$.
When $\Sigma_{\Sel}\succ\Lambda_1^{-1}$, every factor satisfies $\psi(\mu_i^{\Sel})>1$, and hence $\kappa_0^{\Sel}>1$.

\section{Proof of Corollary~\ref{cor:error}}\label{app:error}
\begin{proof}
The outline is to bound the covariance matrix of the stochastic noise $\varepsilon_M(k)$ from below by the constant matrix $\Sigma_\varepsilon/M$, defined in \eqref{eq:Sigma_eps}, and to compare the resulting covariance recursion of the error $\chi(k)$ with the Lyapunov recursion, defined in Step 3.

\textit{Step 1: Error dynamics.}
Subtracting \eqref{eq:nominal} from \eqref{eq:stoch} gives
\begin{equation}\label{eq:chi_dyn}
\chi(k+1)=\bigl(A-BK(\Sel)\bigr)\chi(k)+B\varepsilon_M(k),\quad \chi(0)=0.
\end{equation}
Since the state $x_{\mathrm{nom}}(k)$ of the infinite-sample limit is deterministic, the error $\chi(k)$ is determined by the state $x(k)$.
Hence, using \eqref{eq:eps_mom}, in which the mean of $\varepsilon_M(k)$ for fixed $x(k)$ and $\hat\varpi_{\Sel}(k)$ is zero,
\begin{equation*}
\E[\varepsilon_M(k)]=0,\quad \E[\chi(k)\varepsilon_M(k)^\top]=0.
\end{equation*}
Induction on $k$ using \eqref{eq:chi_dyn} and $\chi(0)=0$ gives $\E[\chi(k)]=0$ and
\begin{equation}\label{eq:cov_prop}
\begin{split}
\mathrm{Cov}[\chi(k+1)]&=\bigl(A-BK(\Sel)\bigr)\mathrm{Cov}[\chi(k)]\bigl(A-BK(\Sel)\bigr)^\top\\
&\quad+B\,\mathrm{Cov}[\varepsilon_M(k)]\,B^\top.
\end{split}
\end{equation}

\textit{Step 2: Lower bound on the noise covariance.}
We bound $\mathrm{Cov}[\varepsilon_M(k)]$ from below.
Using \eqref{eq:kappa} and $\Psi\succ0$, we have $\kappa^{\Sel}\ge\kappa_0^{\Sel}$.
Since $(m_2-m_1)(m_2-m_1)^\top\succeq0$, using \eqref{eq:Xi} gives
\begin{equation*}
\Xi(\kappa^{\Sel})\succeq\kappa^{\Sel}\Lambda_2^{-1}\succeq\kappa_0^{\Sel}\Lambda_2^{-1}.
\end{equation*}
Hence the covariance matrix $P_{\Sel}(0)\Xi(\kappa^{\Sel})P_{\Sel}(0)^\top/M$ of $\varepsilon_M(k)$ for fixed $x(k)$ and $\hat\varpi_{\Sel}(k)$ is no smaller than $\Sigma_\varepsilon/M$ for any $x(k)$ and $\hat\varpi_{\Sel}(k)$.
Since the mean for fixed $x(k)$ and $\hat\varpi_{\Sel}(k)$ is zero, $\mathrm{Cov}[\varepsilon_M(k)]$ equals the expectation of this covariance matrix over $x(k)$ and $\hat\varpi_{\Sel}(k)$, and therefore
\begin{equation*}
\mathrm{Cov}[\varepsilon_M(k)]\succeq\frac{\Sigma_\varepsilon}{M}.
\end{equation*}

\textit{Step 3: Time-wise lower bound.}
Let $\Upsilon(k)$ be given by
\begin{equation}\label{eq:Lyap_rec}
\Upsilon(k+1)=\bigl(A-BK(\Sel)\bigr)\Upsilon(k)\bigl(A-BK(\Sel)\bigr)^\top+B\Sigma_\varepsilon B^\top,
\end{equation}
where $\Upsilon(0)=0$.
For symmetric matrices $X\succeq Y$, $(A-BK(\Sel))X(A-BK(\Sel))^\top\succeq(A-BK(\Sel))Y(A-BK(\Sel))^\top$ holds.
Hence induction on $k$ using \eqref{eq:cov_prop}, \eqref{eq:Lyap_rec}, and $\mathrm{Cov}[\chi(0)]=\Upsilon(0)=0$ gives $\mathrm{Cov}[\chi(k)]\succeq\Upsilon(k)/M$ for all $k\ge0$.
Since $\E[\chi(k)]=0$, we obtain $\E[\|\chi(k)\|^2]=\mathrm{tr}\,\mathrm{Cov}[\chi(k)]\ge\mathrm{tr}\,\Upsilon(k)/M$.

\textit{Step 4: Limit.}
Solving \eqref{eq:Lyap_rec} gives
\begin{equation*}
\Upsilon(k)=\sum_{j=0}^{k-1}\bigl(A-BK(\Sel)\bigr)^jB\Sigma_\varepsilon B^\top\bigl((A-BK(\Sel))^j\bigr)^\top.
\end{equation*}
Under the stability condition \eqref{eq:rho}, this series converges as $k\to\infty$, and its limit is the unique solution $\Upsilon$ of \eqref{eq:Lyap_eq}~\cite{Anderson1979}.
If $\E[\|\chi(k)\|^2]\le\chi_{\mathrm{tol}}^2$ holds for all $k\ge0$, then Step 3 gives $\mathrm{tr}\,\Upsilon(k)\le M\chi_{\mathrm{tol}}^2$ for all $k\ge0$, and letting $k\to\infty$ yields \eqref{eq:Mreq}.
\end{proof}

\section{Proof of Theorem~\ref{thm:kappa0}}\label{app:proof_thm_kappa0}
\begin{proof}
\textit{1)}
The outline is to rewrite the two determinants in \eqref{eq:kappa0} as products of the eigenvalues $\mu_i(N,\Sel)$.

Since
\begin{equation*}
\Sigma_{\Sel}^{-1/2}\bigl(\Sigma_{\Sel}H_{\Sel}\bigr)\Sigma_{\Sel}^{1/2}=\Sigma_{\Sel}^{1/2}H_{\Sel}\Sigma_{\Sel}^{1/2},
\end{equation*}
the matrix $\Sigma_{\Sel}H_{\Sel}$ is similar to $\Sigma_{\Sel}^{1/2}H_{\Sel}\Sigma_{\Sel}^{1/2}$ and has the eigenvalues $\mu_i(N,\Sel)$.
Hence, for a constant $c>0$, the eigenvalues of $I_{md}+c\Sigma_{\Sel}H_{\Sel}$ are $1+c\mu_i(N,\Sel)$, and
\begin{equation*}
\bigl|I_{md}+c\Sigma_{\Sel}H_{\Sel}\bigr|=\prod_{i=1}^{md}\bigl(1+c\mu_i(N,\Sel)\bigr).
\end{equation*}
Applying this expression with $c=2/\lambda$ and $c=4/\lambda$ to the numerator and the denominator of \eqref{eq:kappa0}, respectively, yields \eqref{eq:kappa0_prod}.

\textit{2)}
The outline is to regard the pair $(N,\Sel)$ as a selection from the $T$ candidate basis functions,
to compare the eigenvalues of the two pairs by the Courant--Fischer theorem, and to apply the monotonicity of $\psi$ to \eqref{eq:kappa0_prod}.

\textit{Step 1: Embedding into the full basis.}
Let $L_T(k)$ denote $L(k)$ with $N=T$.
Since the basis function $l_i(k)$, defined in \eqref{eq:li}, does not depend on $N$, Assumption~\ref{assumption:TN} gives $L(k)=[I_N\ 0]L_T(k)$.
Let
\begin{equation*}
\bar\Sel:=\Bigl(I_m\otimes\begin{bmatrix}I_N\\0\end{bmatrix}\Bigr)\Sel\in\{0,1\}^{mT\times md},
\end{equation*}
whose columns are distinct unit vectors.
Using the mixed-product property of the Kronecker product, $P_{\Sel}(k)$, defined in \eqref{eq:vexp}, satisfies
\begin{equation}\label{eq:P_embed}
P_{\Sel}(k)=(I_m\otimes L_T(k)^\top)\bar\Sel.
\end{equation}

\textit{Step 2: Relation between the matrices.}
Let $H_I$ and $\Sigma_I$ denote $H_{\Sel}$ and $\Sigma_{\Sel}$ with $N=T$ and $\Sel=I_{mT}$.
Using \eqref{eq:Bbar} and \eqref{eq:vexp}, we have $\bar B_k\Sel=\sum_{j=0}^{k-1}A^{k-1-j}BP_{\Sel}(j)$.
Substituting \eqref{eq:P_embed} into this expression, \eqref{eq:H}, \eqref{eq:HS_def}, and \eqref{eq:SigmaS} yields
\begin{equation}\label{eq:restrict}
H_{\Sel}=\bar\Sel^\top H_I\bar\Sel,\quad \Sigma_{\Sel}^{-1}=\bar\Sel^\top\Sigma_I^{-1}\bar\Sel.
\end{equation}

\textit{Step 3: Max-min representation of the eigenvalues.}
Applying the Courant--Fischer theorem~\cite{HornJohnson2013} to $\Sigma_{\Sel}^{1/2}H_{\Sel}\Sigma_{\Sel}^{1/2}$ and changing the variables by $\vartheta_{\Sel}=\Sigma_{\Sel}^{1/2}y$ yield
\begin{equation*}
\mu_i(N,\Sel)=\max_{\dim\mathcal{X}=i}\ \min_{\vartheta_{\Sel}\in\mathcal{X}\setminus\{0\}}\frac{\vartheta_{\Sel}^\top H_{\Sel}\vartheta_{\Sel}}{\vartheta_{\Sel}^\top\Sigma_{\Sel}^{-1}\vartheta_{\Sel}},
\end{equation*}
where $\mathcal{X}$ ranges over the $i$-dimensional subspaces of $\mathbb{R}^{md}$.
For $\vartheta\in\mathbb{R}^{mT}\setminus\{0\}$, let
\begin{equation*}
R(\vartheta):=\frac{\vartheta^\top H_I\vartheta}{\vartheta^\top\Sigma_I^{-1}\vartheta}.
\end{equation*}
By \eqref{eq:restrict}, the quotient above equals $R(\bar\Sel\vartheta_{\Sel})$.
Since $\bar\Sel$ is of full column rank, the map $\mathcal{X}\mapsto\bar\Sel\mathcal{X}$ is a one-to-one correspondence between the $i$-dimensional subspaces of $\mathbb{R}^{md}$ and those contained in $\mathrm{Im}(\bar\Sel)$.
Hence
\begin{align*}
\mu_i(N,\Sel)&=\max_{\mathcal{W}\subset\mathrm{Im}(\bar\Sel),\ \dim\mathcal{W}=i}\ \min_{\vartheta\in\mathcal{W}\setminus\{0\}}R(\vartheta),\\
\mu_i(T,I_{mT})&=\max_{\mathcal{W}\subset\mathbb{R}^{mT},\ \dim\mathcal{W}=i}\ \min_{\vartheta\in\mathcal{W}\setminus\{0\}}R(\vartheta).
\end{align*}
The range of the first maximization is included in that of the second, and therefore
\begin{equation*}
\mu_i(N,\Sel)\le\mu_i(T,I_{mT}),\quad i\in\{1,\ldots,md\}.
\end{equation*}

\textit{Step 4: Properties of $\psi$.}
For $\mu\ge0$, the inequality $(1+2\mu/\lambda)^2\ge1+4\mu/\lambda$ gives $\psi(\mu)\ge1$.
In addition, for $\mu\ge0$,
\begin{equation*}
\frac{d}{d\mu}\ln\psi(\mu)=\frac{2}{\lambda+2\mu}-\frac{2}{\lambda+4\mu}\ge0,
\end{equation*}
and hence $\psi$ is monotonically nondecreasing on $\mu\ge0$.

\textit{Step 5: Comparison.}
Since $d\le N\le T$, the dimensions satisfy $md\le mT$.
Using \eqref{eq:kappa0_prod}, Step 3, and Step 4, we obtain
\begin{equation*}
\begin{split}
\kappa_0(N,\Sel)&=\prod_{i=1}^{md}\psi\bigl(\mu_i(N,\Sel)\bigr)\le\prod_{i=1}^{md}\psi\bigl(\mu_i(T,I_{mT})\bigr)\\
&\le\prod_{i=1}^{mT}\psi\bigl(\mu_i(T,I_{mT})\bigr)=\kappa_0(T,I_{mT}),
\end{split}
\end{equation*}
where the first inequality follows from the monotonicity of $\psi$ and the second from $\psi\ge1$.
\end{proof}
\end{document}